\documentclass[eqsecnum,english,prd,aps,nofootinbib,superscriptaddress,amsfonts,longbibliography]{revtex4-1}
\usepackage[T1]{fontenc}
\usepackage[utf8]{inputenc}
\usepackage{amsmath,amssymb,latexsym}
\usepackage{mathrsfs}
\usepackage[unicode=true, pdfusetitle, bookmarks=true,bookmarksnumbered=false,bookmarksopen=false,breaklinks=false,pdfborder={0 0 1},backref=false,colorlinks=false]
 {hyperref}
\usepackage{color}
\usepackage{natbib}
\newtheorem{theo}{Theorem}[section]
\newtheorem{defi}[theo]{Definition}
\newtheorem{bem}[theo]{Remark}
\newtheorem{lemma}[theo]{Lemma}
\newtheorem{koro}[theo]{Corollary}

\newtheorem{conclusion}[theo]{Conclusion}
\newtheorem{ob}[theo]{Observation}
\newtheorem{conjecture}[theo]{Conjecture}
\newtheorem{propo}[theo]{Proposition}

\newcommand{\tit}{\textit}

\newcommand{\R}{\mathbb{R}}
\newcommand{\Z}{\mathbb{Z}}
\newcommand{\beq}{\begin{equation}}
\newcommand{\eeq}{\end{equation}}

\makeatletter

 \@ifundefined{textcolor}{}
 {
   \definecolor{BLACK}{gray}{0}
   \definecolor{WHITE}{gray}{1}
   \definecolor{RED}{rgb}{1,0,0}
   \definecolor{GREEN}{rgb}{0,1,0}
   \definecolor{BLUE}{rgb}{0,0,1}
   \definecolor{CYAN}{cmyk}{1,0,0,0}
   \definecolor{MAGENTA}{cmyk}{0,1,0,0}
   \definecolor{YELLOW}{cmyk}{0,0,1,0}
 }

\makeatother

\allowdisplaybreaks[1]

\begin{document}

\title{The Structurally Dynamic Cellular Network and Quantum Graphity Approaches to Quantum Gravity and Quantum Geometry - A Review and Comparison}

\author{Manfred Requardt}

\email{requardt@theorie.physik.uni-goettingen.de}

\affiliation{Institut fuer Theoretische Physik Universitaet Goettingen\\Friedrich-Hund-Platz 1 37077 Goettingen Germany}

\author{Saeed Rastgoo}

\email{saeed@xanum.uam.mx}

\affiliation{Departamento de F\'{i}sica, Universidad Aut\'{o}noma Metropolitana - Iztapalapa\\ San Rafael Atlixco 186, Mexico D.F. 09340, Mexico}

\date{\today}


\begin{abstract}
Starting from the working hypothesis that both physics and the corresponding mathematics and in particular geometry have to be described by means of discrete concepts on the Planck-scale, one of the many problems one has to face in this enterprise is to find the discrete protoforms of the building blocks of our ordinary continuum physics and mathematics living on a smooth background, and perhaps more importantly find a way how this continuum limit emerges from the mentioned discrete structure. We model this underlying substratum as a structurally dynamic cellular network (basically a generalisation of a cellular automaton).
We regard these continuum concepts and continuum spacetime in particular as being emergent, coarse-grained and derived relative to this underlying erratic and disordered microscopic substratum, which we would like to call quantum geometry and which is expected to play by quite different rules, namely generalized cellular automaton rules. A central role in our analysis is played by a geometric renormalization group which creates (among other things) a kind of sparse translocal network of correlations between the points in classical continuous space-time and underlies, in our view, such mysterious phenomena as holography and the black hole entropy-area law. The same point of view holds for quantum theory which we also regard as a low-energy, coarse-grained continuum theory, being emergent from something more fundamental. In this paper we review our approach and compare it to the quantum graphity framework. 
\end{abstract}


\maketitle

\tableofcontents


\section{Introduction}
In the beautiful book \cite{Ila1} the title of chapter 12 reads: ``Is Nature, underneath it all, a CA?''. Such a point of view contrasts as to the basic philosophy with more prominent approaches to quantum gravity (QG) as e.g. \tit{string theory} while there exist both differences and similarities with respect to \tit{loop quantum gravity}. 
There do exist, nevertheless, various colleagues who think that physics and in particular space-time (S-T) itself behaves basically as a (generalized) CA on the primordial (Planck) scale. One can as well call this primordial state \tit{pregeometry} or \tit{quantum geometry}, thus indicating that a S-T in the ordinary sense does not yet exist. In some of our papers we sometimes dubbed this state \tit{quantum space} or QX. This working philosophy typically considers gravity, continuous space-time and/or quantum theory as being \tit{emergent} or \tit{derived} structures, living over a microscopic primordial substratum which presumably plays by quite different rules as compared to the above \tit{coarse grained} concepts  existing on a more macroscopic scale. In the classification of Carlip  these approaches are called type II (\cite{Carlip}).

While some are openly (e.g. \tit{dynamical triangulation}) or implicitly (\tit{spin-networks}) inspired by geometrical ideas of a more regular type, there are others which rely on quite different sources of inspiration. We would like to take the opportunity to develop this other strand of ideas and concepts and exhibit their roots because whereas some of the aforementioned frameworks are looking superficially quite similar at first glance, under closer inspection it becomes clear that they are drawing their ideas from different sources.

To give an example of the latter class we mention the so-called \tit{quantum graphity} approach (see for example \cite{Konopka1},\cite{Konopka2},\cite{Konopka3}) which we want to compare to our framework being introduced in the following and which may be called \tit{Structurally Dynamic Cellular Networks} (SDCN) or automata (SDCA) approach to quantum geometry.

To put it in a nutshell, there exist in our view more or less three subclasses of this type II classification:
\begin{enumerate}
\item Model theories like dynamical triangulation and quantum graphity in which the matter degrees of freedom (DoF) do not play a primary role or evolve more or less independently relative to the geometric DoF.
\item The cellular automaton interpretation of quantum mechanics (most prominently figuring the work of 't Hooft). From the many papers he published we mention only quite a few; the recent review \cite{Hooft1} and some of the earlier ones, e.g. \cite{Hooft2} and \cite{Hooft3}.\\ 
The characteristic of 't Hooft's approach is a relatively immediate (formal) quantization of classical states (called by him \tit{primordial quantization}) and Hilbert space construction while the static geometric background is the typical regular lattice structure of the CA-framework. 
That means, each of the classical states is promoted to a (basis) vector in some Hilbert space, while the important \tit{superposition principle} is more or less decreed to hold by definition.
\item The SDCN approach undertakes to treat the geometric and quantum matter DoF as \tit{coevolving} structures. The central aim is to create a model of the \tit{quantum vacuum} as the carrier of all the emergent higher and derived structures (as was, for example, the working philosophy of Wheeler (cf. \cite{Wheeler1}).\\  
The first published papers developing this working philosophy were \cite{Requ1},\cite{Requ2} and \cite{Requ3}. The important ingredient was that both the matter states (located at the vertices of the network) and the geometric DoF (sitting at the edges or links) coevolve dynamically with links being allowed to be created and/or deleted in this process. This implies that a geometric \tit{unfolding} of the whole network structure becomes possible together with \tit{geometric phase transitions} which may simulate the situation, for example, prevailing in the big bang and/or inflation era. \\
We want to mention another important point of the SDCN-approach which becomes apparent when we undertake to reconstruct continuum S-T via a coarse graining or \tit{renormalization process}  (cf.\cite{Requren},\cite{Requcon}) from some underlying microscopic network substratum. This process leads almost automatically to the expected \tit{near-order structure} of ordinary continuous space-time physics (locality of the physical processes) on one hand, and to a so-called \tit{far-order} structure on the other hand. The latter means that there always do exist a sparse network of \tit{translocal} links between regions of S-T which are far apart with respect to the distance metric of the ordinary continuous space.We expect that these non-local aspects underlie for example the \tit{quantum entanglement}, the \tit{holgraphic principle} and the \tit{black hole entropy-area law} (see e.g. \cite{Requscale},\cite{Requworm},\cite{Requworm2}.
\end{enumerate}

Before going into the technical details of the network approach to quantum space-time physics we think, it is a good idea to make some brief remarks on the previous history of the discrete network approach to quantum gravity and/or quantum theory as this point of view is, perhaps unjustifiably, a little bit brushed under the carpet as a consequence of the claim that for example \tit{string theory} is \tit{the only game in town} while this theory starts from a fully continuous structure and assumes quantum mechanics to hold unaltered all the way down to the Planck scale.

There do in fact exist various discrete approaches to fundamental physics 
in the recent and not so recent past. For example, already Born and Jordan experimented with discrete \tit{difference equations} before the advent of the correct quantum theory in the twenties. For the sake of brevity we want to begin our sketchy historical remarks with the conceptualization of CA in this context of fundamental physics.

A fairly complete discussion of the history can for example be found in the book by Ilachinski (\cite{Ila1}) together with a more exhaustive bibliography. Such ideas have in fact been around for quite some time (cf. \cite{Zuse},\cite{Feynman},\cite{Fredkin},\cite{Finkelstein}). There was even a whole conference on Physics of Computation, held at MIT in 1982, devoted to such aspects (collected in Vol.21 of the Int.J.Theor.Phys.)

What makes CA so fascinating is the observation that various model classes are capable of universal computing (like e.g. the famous \tit{Game of Life} being invented by Conway; see e.g. \cite{Gardner}). The creation of life on a computer was one of the buzz-words of the emerging field of \tit{Artificial Life} (to mention just a few sources, see e.g. \cite{Kauffman},\cite{Langton},\cite{Waldrop}. After all, CA or SDCN have the potential to realize an important meta idea of modern science, i.e., to generate complex behavior, starting from a, at first glance, relatively simple \tit{dynamical (discrete) system} which is governed by \tit{algorithmic} rules. If it is really the case that at the bottom everything is information (in an admittedly vague sense), this approach has its special merits in the competition with the (perhaps better known) rivals mentioned above.   

However, to model the physical universe on some primordial scale like the Planck scale, we have to satisfy a couple of severe constraints which come from the quantum world and the necessity to unite the quantum world with general relativity. After all, we have to implement both quantum entanglement and the emergence of space-time together with its seething sea of vacuum fluctuations. We think this is more than can be expected from a CA living on a static rigid lattice.
\begin{bem}
It should be emphasized that it is not our aim to merely somehow reproduce certain aspects of, say, quantum theory on a computer. What is actually needed is a structural equivalence of the respective network model and, say, \tit{quantum space-time} with all of its intricate properties (cf. our introductory remarks in \cite{Nowotny1} or \cite{Hooft4})
\end{bem}

Let us make a final remark concerning the various mathematical fields which are involved in our enterprise as this variety of connections  is frequently also invoked as a special merit on the side of string theory. To mention just a few fields: advanced graph theory (e.g. clique graphs, random graphs), Connes' noncommutative analysis and geometry, operator theory on discrete metric spaces (graph Laplacians, Dirac operators, their eigenvalues), generalizations of dimensional concepts, leading even to connections with geometric group theory (via the so-called Cayley graphs), Gromov-Haussdorff Limit of irregular spaces, Small World Networks etc.     

We begin our investigation 
in section \ref{network}
by introducing our underlying dynamical network model and the necessary concepts and notions. We then proceed 
in section \ref{diff-op}
with the derivation of various concepts and tools of discrete (functional) analysis and operator theory on irregular metric spaces. In a next step
in section \ref{dim},
we introduce the concept of generalized dimension on such and non-standard spaces. Then follows the ambitious enterprise to define a geometric renormalization process with classical space-time emerging as some coarse grained limit
in section \ref{Ren}.
 
We will then presents some arguments in section \ref{wormhole-sp}, on how these steps will lead to a better understanding of various crucial concepts of modern physics (in particular concerning \tit{quantum gravity}). examples being e.g. the mysterious phenomenon of \tit{holography}, \tit{quantum entanglement}, the \tit{black hole entropy-area law} etc.
In the last two sections, \ref{qgph} and \ref{compare}, we first make a brief review of another bottom-up model of emergent spacetime called quantum graphity and then try to comment on differences and similarities of our model (SDCN) and the quantum graphity, and make some concluding remarks about our model.


\section{The Microscopic SDCN-Substratum}
\label{network}Our networks are defined on general graphs, $G$, with $V(G)$ the set
of its
vertices (sites or nodes) and 
$E(G)$ its set of edges (links or bonds).
\begin{defi}Here are some graph-theoretical notions and concepts (for more details see e.g. \cite{Bollo1}).
\begin{enumerate}
\item We write the simple, or directed labeled graph as $G:=(V,E)$ where $V$ is the 
  countable set of vertices $\{n_i\}$ 
   and $E$ the set of edges.
   The graph is called simple if there do not exist elementary {\em loops} and {\em multiple edges}. Ann elementary loop is an edge which starts and ends on the same vertex (or connect the same vertex to itself). Multiple edges happen when there is more than one edge directly connecting two vertices.(We could of course also discuss more general graphs).
  Furthermore, for simplicity, we assume the graph to
  be connected, i.e. 
  two arbitrary vertices
  are connected 
    by a sequence of consecutive edges called an {\em edge  sequence} or {\em walk}. A
  minimal edge sequence, that is one with 
  each intermediate vertex
   occurring only once, is called a {\em path}. 
\item For convenience we assume the graph to be {\em locally finite}, 
  which means that the {\em vertex  degree}, $v_i$, or the number of edges incident on each vertex $n_i$ is finite.
 Sometimes it is  useful to make the stronger assumption that $v_i$ is globally bounded away from $\infty$.  
\item One can give the edges both an {\em orientation} and a {\em
    direction} (these two slightly different geometric concepts are
    frequently intermixed in the literature). In an {\em undirected} graph 
    the edges $e_{ij}$ correspond to unordered 
      pairs of vertices $\{n_i,n_j\}$
       while in a {\em directed} graph, the edges 
       have a direction represented 
    by an ordered pair of vertices $d_{ij}=(n_i,n_j)$, 
    that is, the edge points from $n_i$ to $n_j$.
    In this work, we adopt the convention that 
    if two vertices
       $n_i,n_k$  are connected by an edge in an unordered graph, we interpret 
    is as follows: There exists a 
    directed edge, $d_{ik}$, pointing from $n_i$ to $n_k$ and a  directed edge, 
    $d_{ki}$, pointing in the opposite direction. In
    an algebraic sense, which will become clear below, we call
    their {\em superposition}
\beq 
b_{ik}:= d_{ik}-d_{ki}= -b_{ki}
\eeq
    the corresponding {\em oriented edge} 
      (for obvious reasons; the directions are fixed while the orientation can 
    change sign). In a sense the above reflects the equivalence of an {\em 
    undirected graph} with a {\em directed multi-graph} having two 
    directed edges 
       pointing in opposite directions for each 
    undirected edge.
  \end{enumerate}
\end{defi}

We now associate states $s_i$ and $J_{ik}$ with the vertices and edges $n_i$ and $e_{ik}$
The local vertex states $s_i$ can assume values in a certain discrete
set. In the examples we have studied, we follow the philosophy that the
network should be allowed to find its typical range of states via the
imposed dynamics. That is, we allow the $s_i$ to vary in principle
over the set $q \cdot \Z$, with $q$ a certain discrete quantum of
information, energy or some other relevant physical quantity.
The edge states
can assume the values
$J_{ik}\in \{-1,0,+1\}$.

Viewed geometrically we associate the states $J_{ik}= +1,-1,0$ with directed
edges pointing from vertex $n_i$ to $n_k$ (in $+1$ case) or in the opposite direction (in $-1$ case) or, in the
case $J_{ik}=0$, with an empty edge. Our network will be updated after each discrete step of the evolution parameter $t\cdot \tau$, $t\in \Z$ (we may call it somewhat sloppily \tit{clock time} as is done for example in computer science) according to the dynamics introduced below. 
\begin{bem} Note that this is exactly the same procedure as in CA.
\end{bem}
We then have as underlying
substratum a clock time dependent \tit{directed} graph, $G_t$.  Our physical idea
is that at each clock time step, an elementary quantum $q$ is transported
along each existing directed edge in the indicated direction between the two vertices which are connected by this edge.

To implement our general working philosophy of mutual interaction of overall
vertex states and network geometry, we now describe some particular network
laws, which we investigated in greater detail in \cite{Nowotny1} together with a lot of numerical simulation and analysis. We
mainly consider two different classes of evolution laws for vertex and edge
states (for reasons of simplicity we choose units so that $q$ is set equal to one): 
\begin{itemize}
\item Network type I
  
    \begin{align}
     s_{i}(t+1)= & s_{i}(t)+\sum_{k}J_{ki}(t)\\
     J_{ik}(t+1)= & \begin{cases}
     \text{sign}(\Delta s_{ik}), & \textrm{If }|\Delta s_{ik}|\geq\lambda_{2}\vee\big(|\Delta s_{ik}|\geq\lambda_{1}\wedge J_{ik}(t)\neq0\big)\\
     0, & \text{otherwise}
     \end{cases}
     \end{align}     
\item Network type II

  \begin{align}
  s_{i}(t+1)= & s_{i}(t)+\sum_{k}J_{ki}(t)\\
  J_{ik}(t+1)= & \begin{cases}
  \text{sign}(\Delta s_{ik}), & \textrm{If }0<|\Delta s_{ik}|<\lambda_{1}\vee\big(0<|\Delta s_{ik}|<\lambda_{2}\wedge J_{ik}(t)\neq0\big)\\
  J_{ik}(t), & \textrm{If }\Delta s_{ik}=0\\
  0, & \text{otherwise}
  \end{cases}
  \end{align} 
\end{itemize}
where $\Delta s_{ik}= s_{i}(t) - s_k(t)$ and $\lambda_2 \geq
\lambda_1\geq 0$ ($\vee,\wedge$ meaning or, and, respectively). We see that in type I, vertices are connected that have very different internal states, leading to large
local fluctuations, while for the type II, vertices with similar
internal states are connected.
\begin{bem}The role of the $\lambda$ parameters is the following. They define kind of a \tit{hysteresis interval} regulating the switching off and on of edges. We hope that they can be tuned such that the network can perform \tit{phase transitions}. We studied the $\lambda$ dependence  
of various network properties in the computer simulations being discussed in \cite{Nowotny1}.
\end{bem}

We proceed by making some remarks in order to put our approach into the
appropriate context.
\begin{bem}~

\begin{enumerate}
\item It is important that, generically, laws as introduced above, do not
  lead to a reversible time evolution, i.e., there will typically be
  \tit{attractors} or \tit{state-cycles} in the total phase space (the overall
  configuration space 
  of the vertex and edge states). 
   On the other hand, there exist strategies (in the context of cellular automata!) to design
  particular \tit{reversible} network laws (cf. e.g. \cite{Toffoli}) which
  are however, typically of second order in clock time. 
   Usually the existence of attractors is considered to be important for \tit{pattern formation}. On
  the other hand, it may suffice that the
  fraction of the
  phase space, occupied by the  system, shrinks in the course of evolution. That is, one has a flow
  into smaller subvolumes of phase space.
\item In the above class of laws, 
  a direct edge-edge interaction 
   is not yet implemented. Note that it would imply a direct nonlinear action of geometry on itself, similar to the interaction of pure gauge fields in gauge theory.
   We are prepared to incorporate such a (possibly
  important) contribution in a next step if it turns out to be
  necessary. 
    In any case there are not so many ways to do this in a
  sensible way. Stated differently, the class of possible, physically
  sensible interactions is perhaps not so large.
\item We would like to emphasize that the 
  (nondynamical) clock-time,
   $t$, should not be confused with the notion of \tit{physical time},
    i.e., the time operationally employed on much coarser scales. The
  latter is rather supposed to be a collective variable and is
  expected (or hoped!) to emerge via a cooperative effect. Clock-time
  may be an \tit{ideal element}, i.e., a notion which comes from
  outside, so to speak, but -- at least for the time being -- we have
  to introduce some mechanism which allows us to label consecutive
  states or describe \tit{change} or evolution.
\end{enumerate}
\end{bem}

We make the following observation
because it is relevant if one follows the
general spirit of modern high energy physics.
\begin{ob}[Gauge Invariance] The above dynamical law depends nowhere on the 
absolute values of the 
vertex charges $s_i(t)$
but only on their relative differences. By the same token, charge is nowhere created or destroyed. We have
\begin{equation}
\Delta(\sum_{i \in I} s_i)=0
\end{equation}
where, for simplicity, we represent the set of sites by their set of
indices $I$, and $\Delta$ denotes the difference between consecutive
clock-time steps. Put differently, we have conservation of the global vertex charge.
To avoid artificial ambiguities we can, e.g., choose a
fixed reference level 
and take the constraint
\begin{equation}
\sum_{i \in I} s_i= 0
\end{equation} 
as our initial condition.
\end{ob}

Summarizing
the main steps of our working philosophy:
\begin{bem}
 \label{rem2.6} Irrespective of the technical details of the dynamical evolution law
  under discussion, the following, in our view crucial,
  principles should be emulated in order to match fundamental requirements
  concerning the capability of {\em emergent} and {\em complex} behavior.
\begin{enumerate}
\item As is the case with, say, gauge theory or general relativity,
  our evolution law on the surmised primordial level should implement
  the mutual interaction of two fundamental substructures. Put
  sloppily: ``{\em geometry}'' acting on ``{\em matter}'' and vice
  versa, where in our context ``{\em geometry}'' is assumed to
  correspond in a loose sense with the local and/or 
  global edge states
   and ``{\em matter}'' with the structure of 
  the vertex states.
 \item By the same token, the alluded {\em self-referential} dynamical
  circuitry of {\em positive feed back structure} is expected to favor a kind of
  {\em undulating behavior} or {\em self-excitation} in contrast to a return
  to some uninteresting `{\em equilibrium state}' 
   as is frequently  the case in systems consisting of a single component which directly
  feeds back on itself. This propensity for the {\em autonomous}
  generation of undulation patterns is in our view an essential
  prerequisite for some form of ``{\em protoquantum behavior}'' which we
  hope to recover on some coarse grained and less primordial level of
  the network dynamics.
\item In the same sense we expect the overall pattern of 
 switched-on and -off edges 
 to generate a kind of ``{\em protogravity}''.
\end{enumerate}
\end{bem}

We want to comment on a particular intriguing result from our numerical simulation performed in \cite{Nowotny1}, that is the phenomenon of \tit{limit cycles}. Because of the finite phase space of the CA (technically it is
infinite, but the vertex states only fill a finite interval of
$\mathbb Z$ due to the nature of the network laws), network states
will eventually repeat, which leads to a limit cycle because of the
memory-less dynamics. We tested for the appearance of such limit
cycles for different network size $n$ (number of vertices) 
and to our surprise, networks of
type I had with very few exceptions extremely short limit cycles of
period $6$. The exceptions we were able to find, had periods of a
multiple of $6$, the longest found (in a network with $n= 810$ vertices) was
$36$. The prevalence of such short limit cycles is still an open
question and beyond this work. We note in this context that already
S. Kauffmann observed such short cycles in his investigation of
switching nets (\cite{Kauffman}, \cite{Kauffman1}) and found it
very amazing. 
\begin{bem} Our computer simulations employed the following initial conditions. We started with a {\em maxmal complete graph}, i.e. each pair of vertices is connected by an undirected edge. The vertex states were chosen from a uniform distribution scattered over some interval of integers (we tried also other distributions but did not find any significantly different results). The initial values of the edge states were chosen from the set $\{+1,-1\}$ with equal weight.
By this choice we wanted to simulate the initial condition prevalent in the {\em big bang era}.
\end{bem}-

These phenomena of short limit cycles are remarkable in the face of the huge accessible phase
spaces of typical models, and points to some hidden ordering tendencies in
these model classes. What is even more startling is that this phenomenon
prevails also in our network type $1$ 
when we introduce a further
element of possible disorder by allowing edges to be dynamically created and
deleted. We formulate the following hypothesis.
\begin{conjecture}We conjecture that this important phenomenon has its roots
  in the self-referential structure (feed-back mechanisms) of many of the
  used model systems.
\end{conjecture}
It is instructive to observe the emergence of such short cycles in very small
models on paper, setting for example $\lambda_1=\lambda_2 =0$, i.e., no
switching-off of edges and taking $n=2,3$ or $4$. Taking, e.g., $n=2$ and
starting from $s_2(0)= s_1(0) \mod 2$, the network will eventually reach a
state $s_1(t_0)= s_2(t_0)$. Without loss of generality we can assume
$s_1(t_0) = s_2(t_0)= 0$ and $J_{12}(t_0)=1$.  This state develops into a
cycle of length 6 as illustrated in table 2a(1) of \cite{Nowotny1}. For $s_1(0)=
s_2(0)+1 \mod 2$ the state eventually becomes $s_1(t_1) = s_2(t_1) +1$,
without loss of generality $s_1(t_1) = 1$, $s_2(t_1)= 0$,- $J_{12}(t_1) = 1$,
resulting in the dynamics in table 2a(2) of \cite{Nowotny1}. Again, the length of the
cycle is $6$. Hence, $6$ is a good candidate for a short cycle length, which
-- of course -- does not explain why such a short length should appear at
all.

The transients (i.e., the clock time interval between some random initial state; cf. \cite{Nowotny1} and the clock time the system arrives on an attractor)
in networks of type I are also rather short and grow slowly
with the network size. On the other hand, networks of type II have much longer limit cycles and transients. Because of
numerical limitations we were only able to determine cycle lengths for small
networks. We observed that the typical transient and cycle
lengths both grow approximately exponentially (cf. table 2b of \cite{Nowotny1}).

\section{Differential and Operator Calculus on Graphs}\label{diff-op}

In the following section we introduce \textit{differential} and \textit{operator
calculus} on graphs. To some extent this topic carries the flavor
of our own ideas (that is, we surmise that not everything we introduce
below can be found in the standard mathematical literature. We note
for example that, as a minor point, the use of matrices instead of
operators is widespread in the literature). A classical text is for
example \cite{Biggs} and a nice more recent source is \cite{Godsil}.
Our own framework can be found in the papers \cite{Requ1}, \cite{Dirac},\cite{Susy}
where more references are given. In a first step we introduce the
\textit{vertex} and \textit{edge Hilbert spaces} for \textit{directed}
and \textit{undirected graphs} (for reasons of mathematical simplicity
we restrict ourselves to \textit{locally finite graphs}; the more
general situation can of course also be dealt with. For a directed
graph we then have \textit{ingoing} and \textit{outgoing} edges at
each vertex. 
\begin{defi}
We denote the in-vertex degree at vertex $x_{i}$ by $v_{i}^{in}$,
the out-vertex degree by $v_{i}^{out}$ and the local vertex degree
by $v_{i}=v_{i}^{in}+v_{i}^{out}$.  
\end{defi}
For such a graph we can introduce two Hilbert spaces: a vertex Hilbert
space, $\mathcal{H}_{v}$, and an edge Hilbert space, $\mathcal{H}_{d}$,
with orthonormal bases being the set of vertices, $x_{i}$, and the
set of existing directed edges, $d_{ij}$.This means that we introduce a formal scalar product on $\mathcal{H}_{v}$
and $\mathcal{H}_{d}$ respectively with
\begin{equation}
\langle x_{i},x_{j}\rangle=\delta_{ij}\quad,\quad\langle d_{ij},d_{lm}\rangle=\delta_{il}\delta_{jm}.
\end{equation}
Then the vectors in these spaces
can be written as the formal sums 
\begin{equation}
f=\sum_{i=1}^{\infty}f_{i}x_{i}\quad,\quad g=\sum_{i,j=1}^{\infty}g_{ij}d_{ij}\;\text{with}\; f_{i},g_{ij}\in\mathbb{C}
\end{equation}
with $\sum|f_{i}|^{2}<\infty$ and $\sum|g_{ik}|^{2}<\infty$.\\[0.3cm] 
\begin{bem}
We treat the vertices and edges as abstract basis elements (in a way
similar to the \textit{group algebra} of a group). One can of course
consider the abstract vectors equally well as discrete functions over
the vertex- or edge-set respectively, and the basis vectors as elementary
indicator functions. Therefore we replace from now on the vertices
$n_{i}$ by the corresponding indicator functions $x_{i}$, having
the value one at the respective vertex under discussion and zero elsewhere. 
\end{bem}
If we deal with an undirected but orientable graph we find it convenient
to introduce the superposition 
\begin{equation}
b_{ij}:=d_{ij}-d_{ji}=-b_{ji}
\end{equation}
and relate it to an undirected but orientable edge. We now introduce
two operators, interpolating between $\mathcal{H}_{v}$ and $\mathcal{H}_{d}$.
We define them on the basis vectors: 
\begin{align}
d: & \mathcal{H}_{v}\to\mathcal{H}_{d}\\
d(x_{i}):= & \sum_{k}d_{ki}-\sum_{k'}d_{ik'}
\end{align}
with the first sum running over the ingoing edges relative to $x_{i}$
and the second sum running over the outgoing ones. In the case of
a symmetric (or undirected graph) we have 
\begin{equation}
d(x_{i}):=\sum_{k}\left(d_{ki}-d_{ik}\right)=\sum_{k}b_{ki}.
\end{equation}
This operator is closely related to a sort of non-commutative discrete
differential calculus on graphs as we have 
\begin{equation}
df=\sum_{i,k}(f_{k}-f_{i})d_{ik}.
\end{equation}
A simple calculation shows that the adjoint, $d^{*}:\mathcal{H}_{d}\to\mathcal{H}_{v}$,
acts on the basis vectors of $\mathcal{H}_{d}$ as follows:
\begin{equation}
d^{*}(d_{ik})=x_{k}-x_{i}
\end{equation}
\begin{bem} Note that these operators are closely related to the {\em boundary} and {\em coboundary operator} in {\em algebraic topology}.
\end{bem}

In algebraic graph theory (finite graphs) the so-called \textit{incidence
matrix}, $B$, is introduced, having the entry $1$ if vertex $x_{i}$
is the positive end of a certain (ingoing) edge, and having a $-1$
if it is the negative end (outgoing edge) (see for example \cite{Godsil}).
This matrix corresponds to our operator $d^{*}$.

Another important operator is the \textit{adjacency matrix}, $A,$
being a map from $\mathcal{H}_{v}$ to $\mathcal{H}_{v}$ and having
(in ordinary graph theory of (un)oriented graphs) a $+1$ at entry
$(i,j)$ if $x_{i}$ and $x_{j}$ are connected by an edge. This matrix
is a symmetric operator, $a_{ij}=a_{ji}$. In our more general context
(which includes however the ordinary situation as a special case)
of directed graphs one can introduce the \textit{in-adjacency} matrix,
$A^{in}$, and the \textit{out-adjacency} matrix, $A^{out}$, with
$A=A^{in}+A^{out}$, $A^{in}$ having a $+1$ at entry $(k,i)$ if $x_k$ and $x_i$ are connected by a directed edge $d_{ki}$ and correspondingly for $A^{out}$. 
\begin{bem}
Note that our $A$ differs slightly from the classical $A$. The classical
$A$ for an undirected graph equals our $A^{in}$ or $A^{out}$ in
that case. Our operators apply to more general situations with our
$A$ being even symmetric for arbitrary directed graphs.  
\end{bem}
In our (operator)-notation they are given by 
\begin{align}
A\, x_{i}=\sum_{k\sim i}\epsilon_{ki}\, x_{k}, &  & A^{in}\, x_{i}=\sum_{k\to i}x_{k}, &  & A^{out}\, x_{i}=\sum_{i\to k'}x_{k'}
\end{align}

with $\sim$ designating the unordered pair $\{x_{i},x_{k}\}$, $k\to i$
the ordered pair $(k,i)$ and $\epsilon_{k,i}$ is either $1$ or
$2$, depending on the two possible cases of one directed edge between
vertex $x_{i}$ and vertex $x_{k}$ or two directed edges, pointing
in opposite directions.

These operators can be built up from more elementary operators (cf.
\cite{Dirac} or \cite{Susy})
 $d_{I}:\mathcal{H}_{v}\to\mathcal{H}_{d},\, d_{I}^{*}:\mathcal{H}_{d}\to\mathcal{H}_{v}$ where $I=1,2$
\begin{align}
d_1\;x_i=\sum_k d_{ki}, &\qquad d_2\;x_i=\sum_{k'} d_{ik'}\\
d_{1}^{*}\; d_{ik}=x_{k}, & \qquad d_{2}^{*}\, d_{ik}=x_{i}\\
d=d_{1}-d_{2}, & \qquad d^{*}=d_{1}^{*}-d_{2}^{*}\\
d_{1}^{*}d_{1}\, x_{i}=v_{i}^{in}\cdot x_{i}, & \qquad d_{2}^{*}d_{2}\, x_{i}=v_{i}^{out}\cdot x_{i}\\
d_{1}^{*}d_{2}\, x_{i}=\sum_{i\to k'}x_{k'}, & \qquad d_{2}^{*}d_{1}\, x_{i}=\sum_{k\to i}x_{k'}
\end{align}
where $v_{i}^{in}\,,\, v_{i}^{out}$ is the in-, out-degree of vertex
$x_{i}$ respectively. We hence have 
\begin{lemma}
The in-, out-vertex degree matrices (having the respective vertex degrees on the diagonal) read 
\begin{equation}
V^{in}=d_{1}^{*}d_{1}\quad,\quad V^{out}=d_{2}^{*}d_{2}.
\end{equation}
The in-, out-adjacency matrices read 
\begin{equation}
A^{in}=d_{2}^{*}d_{1}\quad,\quad A^{out}=d_{1}^{*}d_{2}
\end{equation}
and
$A=A^{in}+A^{out}$ is symmetric.  \end{lemma}
\begin{propo}
The so-called graph Laplacian is the following positive operator  
\begin{equation}
-\Delta:=d^{*}d=\left(V^{in}+V^{out}\right)-\left(A^{in}+A^{out}\right)=V-A.
\end{equation}
\end{propo}
Note that for an undirected graph (i.e. both $d_{ik}$ and $d_{ki}$
being present) the above Laplacian is simply twice the classical Laplacian
(matrix). 

The reason to call this operator a Laplacian stems from the observation
that it acts like a second order partial difference operator on functions
of $\mathcal{H}_{0}$. 
\begin{equation}
-\Delta\, f=\sum_{i}f_{i}\left(v_{i}^{in}x_{i}+v_{i}^{out}x_{i}-\sum_{k\to i}x_{k}-\sum_{i\to k}x_{k}\right)
\end{equation}
and after a simple relabeling of indices 
\begin{align}
-\Delta\, f=&-\sum_{i}\left(\sum_{k\to i}f_{k}+\sum_{i\to k}f_{k}-v_{i}^{in}f_{i}-v_{i}^{out}f_{i}\right)x_{i}\nonumber\\
=&-\sum_{i}\left(\sum_{k\to i}(f_{k}-f_{i})+\sum_{i\to k}(f_{k}-f_{i})\right)x_{i}\nonumber\\
=&-\sum_{i}\left(\sum_{k\sim i}\epsilon_{ki}(f_{k}-f_{i})\right)x_{i}
\end{align}
which reduces to the ordinary expression in the undirected case.

Forming now the direct some $\mathcal{H}:=\mathcal{H}_{v}\oplus\mathcal{H}_{d}$,
we can introduce yet another important graph operator which closely
entangles geometric and functional analytic properties of graphs (and
similar structures); see \cite{Dirac} and \cite{Susy}. 
\begin{defi}
We define the graph Dirac operator as follows 
\begin{align}
D:\,\mathcal{H}\to\mathcal{H} &  & \text{with \,\,\,}D:=\left(\begin{array}{cc}
0 & d^{\ast}\\
d & 0
\end{array}\right)\quad,\quad\mathcal{H}=\left(\begin{array}{c}
\mathcal{H}_{0}\\
\mathcal{H}_{1}
\end{array}\right).
\end{align}
   \end{defi}
\begin{ob}
We have 
\begin{equation}
D^{2}=D\, D=\left(\begin{array}{cc}
d^{\ast}d & 0\\
0 & dd^{\ast}
\end{array}\right)
\end{equation}
with $d^{*}d=-\Delta$. 
\end{ob}
The action of $dd^{*}$ on a basis vector $d_{ik}$ reads 
\begin{equation}
dd^{*}\, d_{ik}=d\,(x_{k}-x_{i})=\sum_{k'}d_{k'k}-\sum_{k''}d_{kk''}-\sum_{l}d_{li}+\sum_{l'}d_{il'}
\end{equation}
which, after some relabeling and introduction of the Kronecker delta
symbol, can be written as 
\begin{equation}
dd^{*}\, d_{ik}=\sum_{m,j}\left(d_{mj}\delta_{jk}-d_{jm}\delta_{jk}-d_{mj}\delta_{ij}+d_{jm}\delta_{ij}\right).
\end{equation}
For a function $g=\sum g_{ik}d_{ik}$ we hence get 
\begin{equation}
dd^{*}\, g=\sum_{l,m}\left(\sum_{i}g_{im}-g_{mi}-g_{il}+g_{li}\right)d_{lm}.
\end{equation}
  There is a pendant in the calculus of differential forms on general
Riemannian manifolds where, with the help of the \textit{Hodge-star
operation}, we can construct a dual, $\delta$, to the ordinary exterior
derivative. The generalized Laplacian then reads 
\begin{equation}
-\Delta=\delta d+d\delta
\end{equation}
with $\delta$ (modulo certain combinatorial prefactors) corresponding
to our $d^{*}$ (see for example \cite{Felsager} or \cite{Hodge}).

With the help of the machinery we have introduced above further properties
of the graph Laplacian and Dirac operator can be derived, some of
them being directly related to geometric and/or wiring properties
of the graph under discussion. Of particular relevance are spectral
properties of $-\Delta$ and $D$. In most of the graph literature
the graphs are assumed to be finite, hence the corresponding operators
are automatically bounded and $-\Delta$ and $D$ are self adjoint.
A fortiori the spectrum is discrete as the corresponding Hilbert spaces
are finite dimensional. This is the reason why most of the graph literature
used the matrix calculus (which is in our view a little bit clumsy).
Investigation of the unbounded case are less numerous. We discussed
this more general case in \cite{Dirac} and \cite{Susy} where also
more references are mentioned. A particular result is for example:
 
\begin{theo}
For a globally bounded vertex degree, $-\Delta$ and $D$ are bounded
self adjoint operators with their bounds being explicitly computable.
(For a proof see \cite{Dirac} and/or \cite{Susy}).) 
\end{theo}
It is noteworthy that with the help of the tools we have introduced
and developed above we can successfully deal with various interesting
modern topics of mathematical physics in this particular context of
irregular discrete spaces. To mention a few cases, in \cite{Requ1},\cite{Dirac},\cite {Susy}
we treated graphs as models of \textit{noncommutative geometry} and
\textit{supersymmetry}. Among other things we introduced an example
of Connes' \textit{spectral triple}, and calculated the so-called
\textit{Connes distance metric} on graphs. That our networks/graphs
carry automatically and naturally a supersymmetric structure may perhaps
be a further hint that this approach and its continuum limit has something
to do with the real physical world of high-energy physics.

\section{Dimensional Concepts on Graphs or Networks}\label{dim}

There exist a variety of concepts in modern mathematics which extend
the ordinary or naive notion of \textit{dimension} one is accustomed
to in e.g. differential topology or linear algebra. In fact, \textit{topological
dimension} and related concepts are notions which are perhaps even
closer to the underlying intuition (cf. e.g. \cite{Edgar}).

Apart from the purely geometric concept, there is also an important
physical role being played by something like dimension, having pronounced
effects on the behavior of, say, many-body-systems near their \textit{phase
transition points} or in the \textit{critical region}.

But even in the case of e.g. lattice systems they are usually treated
as being embedded in an ambient continuous background space (typically
euclidean) which supplies the concept of ordinary dimension so that
the \textit{intrinsic dimension} of the discrete array itself does
not usually openly enter the considerations.

Anyway, it is worthwhile even in this relatively transparent situation
to have a closer look on the situations where attributes of something
like dimension really enter the physical stage. Properties of models
of, say, statistical mechanics are typically derived from the structure
of the microscopic interactions of their constituents. This then is
more or less the only place where dimensional aspects enter the calculations.

Naive reasoning might suggest that it is something like the number
of nearest neighbors (in e.g. lattice systems) which encodes the dimension
of the underlying space and influences via that way the dynamics of
the system. However, this surmise, as we will show in the following,
does not reflect the crucial point which is more subtle.

This holds the more so for systems which cannot be considered as being
embedded in a smooth regular background and hence do not inherit their
dimension from the embedding space. A case in point is our primordial
network in which Planck-scale-physics is assumed to take place. In
our approach, it is in fact exactly the other way round: smooth space-time
is assumed to emerge via a (geometric) \textit{phase transition} or
a certain \textit{cooperative behavior} and after some ``\textit{coarse
graining}'' from this more fundamental structure. That is, our task
is to formulate an intrinsic notion of dimension for model theories
living on quite irregular spaces without making recourse to the dimension
of some continuous embedding space.

In a first step we will show that graphs or networks as introduced
in the preceding sections carry a natural metric structure. We have
already introduced a certain neighborhood structure in a graph with
the help of the minimal number of consecutive edges connecting two
given vertices. In a connected graph any two vertices can be connected
by a sequence of edges. Without loss of generality one can restrict
oneself to \textit{paths}. One can then define the length of a path
(or sequence of edges) by the number $l$ of consecutive edges making
up the path. 
\begin{bem} Unless otherwise stated we deal in the following with undirected graphs, that is we regard our networks or graphs as \tit{metric spaces}.
\end{bem}
\begin{ob}
Among the paths connecting two arbitrary vertices $x_{i},\, x_{k}$
there exists at least one with minimal length which we denote by $d(x_{i},x_{k})$. We apologize for using again the letter $d$ for distance but this is a standard notation for a distance metric. Furthermore, the differential operator $d$ will not appear again in the following so that there should not arise any confusion.
This $d$ has the properties of a metric, i.e: 
\begin{align}
d(x_{i},x_{i})= & 0\\
d(x_{i},x_{k})= & d(x_{k},x_{i})\\
d(x_{i},x_{l})\leq & d(x_{i},x_{k})+d(x_{k},x_{l}).
\end{align}
(The proof is more or less evident). \end{ob}
\begin{koro}
With the help of the metric, one gets a natural neighborhood structure
around any given vertex, where $B_{m}(x_{i})$ comprises all the vertices,
$x_{k}$, with $d(x_{i},x_{k})\leq m$, and $\partial B_{m}(x_{i})$
the vertices with $d(x_{i},x_{k})=m$.
\end{koro}
This natural neighborhood structure enables us to develop the concept
of an intrinsic dimension on graphs and networks. To this, one first
has to realize what property really matters physically (e.g. dynamically),
independent of the kind of model or embedding space. 
\begin{ob}
The crucial and characteristic property of, say, a graph or network
which may be associated with something like dimension is the number
of ``new vertices'' in $B_{m+1}$ compared to $B_{m}$ for $m$
sufficiently large or $m\to\infty$. The deeper meaning of this quantity
is that it measures the kind of ``wiring'' or ``connectivity''
in the network and is therefore a ``topological invariant''.
\end{ob}
Regarding the graph as an example of a metric space we shall replace
the discrete value $m$ by the continuous parameter $r$, hence writing
the ball-neighborhoods as $B(x,r)$. We then define the \textit{growth
function} and \textit{spherical growth function} on $G$ relative
to some arbitrary but fixed vertex $x$. (We use here the notion more
common in e.g. geometric group theory. In other fields it is also
called the distance degree sequence, cf. \cite{Harary}). 
\begin{defi}
The growth function $\beta(G,x,r)$ is defined by 
\begin{equation}
\beta(G,x,r)=|B(x,r)|
\end{equation}
with $|B(x,r)|$ denoting the number of vertices in $B(x,r)$.\\
Correspondingly we define 
\begin{equation}
\partial\beta(G,x,k):=\beta(G,x,k)-\beta(G,x,k-1).
\end{equation}

\end{defi}
With the help of the limiting behavior of $\beta$ and $\partial\beta$,
we introduce two dimensional concepts. 
\begin{defi}
The (upper, lower) internal scaling dimension with respect to the
vertex $x$ is given by 
\begin{align}
\overline{D}_{s}(x):=\limsup_{r\to\infty}(\ln\beta(x,r)/\ln r), &  & \underline{D}_{s}(x):=\liminf_{r\to\infty}(\ln\beta(x,r)/\ln r)
\end{align}

The (upper,lower) connectivity dimension is defined correspondingly
as 
\begin{align}
\overline{D}_{c}(x):=\limsup_{k\to\infty}(\ln\partial\beta(x,k)/\ln k)+1, &  & \underline{D}_{k}(x):=\liminf_{k\to\infty}(\ln\beta(x,k)/\ln k)+1.
\end{align}

If the upper and lower limits coincide, we call it the internal scaling
dimension, the connectivity dimension, respectively.  \end{defi}
\begin{bem}
~
\begin{enumerate}
\item The two notions are not entirely the same in general, whereas they
coincide for many models (this is quite similar to the many different
fractal dimensions). 
\item For regular lattices, both concepts
yield the expected result, i.e. the embedding dimension. In general
however, the upper and lower limits are different and non-integer.
Similarities to fractal dimensions are not accidental. For a more
thorough discussion of all these points see \cite{Requ3}.
\end{enumerate}
\end{bem}
Some historical remarks are perhaps in order. We developed and investigated
this concepts in \cite{Requ3} almost from scratch. We later observed
that there existed some scattered remarks in the literature using
similar concepts but, as to our knowledge, nowhere were the extremely
interesting properties of this concept studied in detail in the physics
literature (cf. the remarks and references in \cite{Requ3}, some ideas can also be found in \cite{Ila1}). On the
other hand we later (roughly at the time of writing \cite{Requcon})
came upon similar concepts employed in a completely different context,
that is, a field of pure mathematics called \textit{geometric group
theory} (see e.g. \cite{Harpe}). We come back to this point below
when we introduce the renormalization and coarse graining process
on graphs or networks. 

It is important that these notions display a marked rigidity against
all sorts of deformations of the underlying graph and are independent
of the reference vertex for locally finite graphs. We mention only
two properties in this direction. 
\begin{ob}
If the vertex-degree of the graph is locally finite, the numerical
values of the above quantities are independent of the reference vertex.\\
(The simple proof can be found in \cite{Requ3}). 
\end{ob}
In the following theorem we prove stability of the graph dimension
under local perturbations of the wiring of graphs. In the first step
we add edges in the $k$-neighborhoods of each vertex. In the second
part of the theorem the local deformations are slightly more complicated. 
\begin{defi}
We pass over from a graph $G$ to a new graph $G'$, having the same vertex set, by means of a number of edge deletions. These edge deletions are
called local of order $k$ if only edges between pairs of vertices $x,\,y$
are deleted which have a distance $d(x,\,y)$ in $G'$ being globally bounded by $k$
(note that the distance metrics in $G$ and $G'$ will differ in general!).
\end{defi}
\begin{theo}
With $G$ locally finite the following holds: 
\begin{enumerate}
\item Insertions of arbitrarily many edges within a $k$-neighborhood of
any vertex do not alter the above dimensional notions.\\
(The proof and a slight generalization can be found in Lemma 4.10
of \cite{Requ3}). 
\item Edge deletions fulfilling the above described property also do not
alter the dimensional notions on the graph.\\ 
(This can be proved by reversing the process, i.e., we pass from the
graph $G'$ to $G$ by $k$-local edge insertions; cf. Theorem 6.8
in \cite{Requworm}).
\end{enumerate}
\end{theo}
Such deformation results are very useful because it turns out to be
surprisingly difficult to construct sufficiently irregular large graphs
with presrcibed properties, for example, having a prescribed (possibly
noninteger) dimension. The above theorem guarantees that irregular
graphs which can be constructed via appropriate deformations starting
from e.g. regular graphs will have the same dimension. For more results
in this direction see the following sections (or \cite{Requ3} and
\cite{Requcon}). 

We want to conclude this section with a remark concerning the nature
of the above defined graph dimension. At a first glance, it may remind
the reader of the various \textit{fractal dimensions} (see e.g. \cite{Falconer})
but this impression is not entirely correct. In a sense it is just
the opposite of a fractal dimension. While fractal dimension is related
to the infinitesimal structure of (irregular) sets, it is in our case
the large distances which matter. Therefore the notion of \textit{growth
degree} is a better description. The reason for this duality stems
from our working philosophy to construct the continuum from some discrete
irregular underlying structure by performing a continuum limit which includes as an essential ingredient a \tit{rescaling process}.
This will become clearer in the following and is just the opposite of going into the infinitesimal small as in fractal geometry. 

It is a characteristic of our construction that we go to large distances
on the underlying graphs (as is the case in the lattice approximations in quantum field theory). After all, to arrive at a rigorous definition,
the graphs have to be infinite. For large but finite graphs we can
of course use the concept in an approximate way. Going to large and
at the end infinite distances is also crucial when we take the continuum
limit in order to reconstruct a corresponding continuum theory. In
this process we rescale the original graph distance metric, that is
we go over from the original distance $d(x,y)$ to $\lambda\cdot d(x,y)$
with $\lambda\to0$. Consequently points which lie very far apart
in the underlying graph or network $G$ become infinitesimal neighbors
in the continuum limit. That is, the growth degree characterizes in
the end the infinitesimal neighborhood of points in the continuum
which is a property of the notion of dimension in the continuum.

As a last remark we want to mention another concept of dimension which is frequently employed in the physics of statistical and critical systems on irregular geometric structures. It is called the \tit{spectral dimension}. As far as we are aware of, early attempts can be found in \cite{Dhar}. A careful mathematical analysis is made in \cite{Watanabe}. A nice paper, relating the spectral dimension to our scaling dimension, is \cite{Filk}. Another investigation of these dimensions is made in \cite{Durhuus} (we emphasize that our list is quite incomplete). The spectral dimension is closely related to diffusion processes on the underlying networks and the \tit{return probability} of \tit{random walks}. Both types of dimensions encode certain geometric properties of the network which can be associated with something like
``geometric'' or ``physical''?
dimension. There are some connections between these notions but generically they are numerically different.


\section{Dynamical Networks as Random Graphs}\label{sec:Random-Networks}

\subsection{The Statistical Hypothesis}

As we are dealing with very large graphs, which are a fortiori constantly
changing their shape, that is, their distribution of (active) edges,
we expect the dynamics to be sufficiently stochastic so that a point
of view may be appropriate which reminds of the working philosophy
of \textit{statistical mechanics}. This, however, does not imply that
our evolving network is nothing but a simple \textit{random graph}
as introduced below (cf. the remarks at the end of this section).
It rather means that some of its geometric characteristics can, or
should, be studied within this well-developed context.

Visualizing the characteristics and patterns being prevalent in large
and ``typical'' graphs was already a notorious problem in \textit{combinatorial
graph theory} and led to the invention of the \textit{random graph}
framework (see the more complete discussion in \cite{lumps}). The
guiding idea is to deal with graphs of a certain type in a probabilistic
sense. This turns out to be particularly fruitful as many graph characteristics
(or their absence) tend to occur with almost certainty in a probabilistic
sense (as has been first observed by Erd\"os and R\'enyi). The standard
source is \cite{Bollo2} (for further references see \cite{lumps}).

Another strand of ideas stems from the theory of dynamical systems
and cellular automata, where corresponding statistical and ensemble
concepts are regularly employed. Typically, we are looking for \textit{attractors}
in phase space, which are assumed to correspond to large scale (that
is, after \textit{coarse graining} and \textit{rescaling}), quasi continuous
or macroscopic patterns of the system. Experience shows, that such
a structure or the approach towards attractors is in many cases relatively
robust to the choice of initial configurations or microscopic details
and, hence, suggests an ensemble picture.

Furthermore, since the early days of statistical mechanics, the ensemble
point of view is, at least partly, corroborated by the philosophy
that time averages transform (under favorable conditions) into ensemble
averages. In our context this means the following. Denoting the typical
length/time scale of ordinary quantum theory by $[l_{qm}],[t_{qm}]$,
we have 
\begin{equation}
[l_{qm}]\gg[l_{pl}]\quad,\quad[t_{qm}]\gg[t_{pl}]
\end{equation}
where the latter symbols denote the Planck scale. Under renormalization,
the mesoscopic scales comprise a huge number of microscopic clock
time intervals and degrees of freedom of the network under discussion.

A fortiori, the networks we are interested in correspond to graphs
having a large \textit{vertex degree}, i.e. channels entering a given
typical vertex of the graph. That is, we expect large local fluctuations
in microscopic grains of space or time. Put differently, the network
locally traverses a large number of different microscopic states in
a typical mesoscopic time interval, $[t_{qm}]$. This observation
suggests that on a mesoscopic or macroscopic scale, microscopic patterns
will be washed out or averaged over.
\subsection{The Random Graph Framework}
\label{random}One kind of probability space is constructed as follows. Take all
possible labeled graphs over $n$ vertices as probability space $\mathcal{G}$
(i.e. each graph represents an elementary event). The maximal possible
number of edges is $N:=\binom{n}{2}$, which corresponds to the unique
\emph{simplex graph} (denoted usually by $K_{n}$). Give each edge
the \emph{independent probability} $0\leq p\leq1$, (more precisely,
$p$ is the probability that there is an edge between the two vertices
under discussion). Let $G_{m}$ be a graph over the above vertex set,
$V$, having $m$ edges. Its probability is then 
\begin{equation}
pr(G_{m})=p^{m}\cdot q^{N-m},
\end{equation}
where $q:=1-p$. There exist $\binom{N}{m}$ different labeled graphs
$G_{m}$, having $m$ edges, and the above probability is correctly
normalized, i.e. 
\begin{equation}
pr(\mathcal{G})=\sum_{m=0}^{N}\binom{N}{m}p^{m}q^{N-m}=(p+q)^{N}=1.
\end{equation}
This probability space is sometimes called the space of  \emph{binomially
random graphs} and denoted by $\mathcal{G}(n,p)$. Note that the number
of edges is binomially distributed, i.e. 
\begin{equation}
pr(m)=\binom{N}{m}p^{m}q^{N-m}
\end{equation}
and 
\begin{equation}
\langle m\rangle=\sum m\cdot pr(m)=N\cdot p.
\end{equation}

The really fundamental observation made already by Erd\"os and R\'enyi
(a rigorous proof of this deep result can e.g. be found in \cite{Bollo2})
is that there are what physicists would call \textit{phase transitions}
in these \textit{random graphs}. To go a little bit more into the
details we have to introduce some more graph concepts. 
\begin{defi}[Graph Properties]
\emph{ Graph properties} are certain particular \emph{random variables}
(indicator functions of so-called events) on the above probability
space $\mathcal{G}$. i.e., a graph property, $Q$, is represented
by the subset of graphs of the sample space having the property under
discussion.  
\end{defi}
To give some examples: i) connectedness of the graph, ii) existence
and number of certain particular subgraphs (such as subsimplices etc.),
iii) other geometric or topological graph properties etc.

In this context Erd\"os and R\'enyi made the following important
observation: 
\begin{ob}[Threshold Function]
A large class of \emph{graph properties} (e.g. the \emph{monotone
increasing ones}, cf. \cite{Bollo2} or \cite{Bollo3}) have a so-called
\emph{threshold function}, $p^{*}(n)$ or $m^{*}(n)$, with $m^{*}(n):=N\cdot p^{*}(n)$, $p^{*}(n)$ a $n$-dependent edge probability,
so that for $n\to\infty$ the graphs under discussion have \emph{property}
$Q$ \emph{almost surely} for $p(n)>p^{*}(n)$ and \emph{almost surely
not} for $p(n)<m^{*}(n)$ or vice versa (for the details see the above cited literature). That is, by turning
on the probability $p$, one can drive the graph one is interested,
in beyond the phase transition threshold belonging to the graph property
under study. Note that by definition, threshold functions are only
unique up to ``factorization'', i.e. $m_{2}^{*}(n)=O(m_{1}^{*}(n))$
is also a threshold function.
\end{ob}
Calculating these graph properties is both a fascinating and quite
intricate enterprise. In our context we are mainly interested in properties of \textit{cliques} (defined below)
, their distribution (with respect to their order, $r$, i.e. number
of vertices), frequency of occurrence of cliques of order $r$, degree
of mutual overlap etc. (cf. also \cite{lumps} and \cite{Requ3}).
These cliques shall be the building blocks of our geometric renormalization
process being described in the following section. We shall relate
these properties to the various assumed stages and phases of our space-time
manifold. 
\begin{defi}
A subsimplex is a subgraph with all its vertices being connected to
each other. A cliques is a maximal subsimplex, i.e., adding another
vertex to the subsimplex will destroy the property of being a subsimplex. 
\end{defi}
We can introduce various \textit{random function} on the above probability
space. For each subset $V_{i}\subset V$ of order $r$, we define
the following random variable: 
\begin{equation}
X_{i}(G):=\begin{cases}
1 & \text{if \ensuremath{G_{i}} is an \ensuremath{r}-simplex},\\
0 & \text{else}
\end{cases}
\end{equation}
where $G_{i}$ is the corresponding induced subgraph over $V_{i}$
in $G\in{\cal G}$ (the probability space). Another random variable
is then the \textit{number of $r$-simplices in $G$}, denoted by
$Y_{r}(G)$ and we have: 
\begin{equation}
Y_{r}=\sum_{i=1}^{\binom{n}{r}}X_{i}
\end{equation}
with $\binom{n}{r}$ the number of $r$-subsets $V_{i}\subset V$.
With respect to the probability measure introduced above we have for
the \textit{expectation values}: 
\begin{equation}
\langle Y_{r}\rangle=\sum_{i}\langle X_{i}\rangle
\end{equation}
and 
\begin{equation}
\langle X_{i}\rangle=\sum_{G\in\mathcal{G}}X_{i}(G)\cdot pr(G_{i}=r\text{-simplex in}\; G).
\end{equation}
These expectation values were calculated in \cite{lumps}. We have
for example 
\begin{equation}
\langle X_{i}\rangle=p^{\binom{r}{2}}.
\end{equation}

The probability that such a subsimplex is maximal, i.e. is a cliques,
is then (cf. \cite{lumps}) 
\begin{equation}
pr(G_{r}\;\text{is a clique})=(1-p^{r})^{n-r}\cdot p^{\binom{r}{2}}.
\end{equation}
As there exist exactly $\binom{n}{r}$ possible different $r$-subsets
in the vertex set $V$, we arrive at the following conclusion: 
\begin{conclusion}[Distribution of Subsimplices and Cliques]
The expectation value of the random variable ``\emph{number of $r$-subsimplices}''
is 
\begin{equation}
\langle Y_{r}\rangle=\binom{n}{r}\cdot p^{\binom{r}{2}}.
\end{equation}
For $Z_{r}$, the \emph{number of $r$-cliques} (i.e. maximal! $r$-simplices)
in the random graph, we have then the following relation 
\begin{equation}
\langle Z_{r}\rangle=\binom{n}{r}\cdot(1-p^{r})^{n-r}\cdot p^{\binom{r}{2}}.
\end{equation}
\end{conclusion}
These quantities, as functions of $r$ (the \textit{order} of the
subsimplices) have quite a peculiar numerical behavior. We are interested
in the typical \textit{order of cliques} occurring in a generic random
graph (where typical is understood in a probabilistic sense. 
\begin{defi}[Clique Number]
The maximal order of occurring cliques contained in $G$ is called
its \emph{clique number}, $cl(G)$. It is another random variable
on the probability space $\mathcal{G}(n,p)$.  
\end{defi}
It is remarkable that this value is very sharply defined in a typical
random graph. Using the above formula for $\langle Z_{r}\rangle$,
we can give an approximate
value, $r_{0}$, for this expectation value and get 
\begin{equation}
r_{0}\approx\frac{2\log(n)}{\log\left(p^{-1}\right)}+O\left(\log\left(\log(n)\right)\right)
\end{equation}
(cf. chapt. XI.1 of \cite{Bollo2}). It holds that practically all
the occurring cliques fall in the interval $(r_{0}/2,r_{0})$. We
illustrate this with the following tables. Our choice for $n$, the
number of vertices, is $10^{100}$. The reason for this seemingly
very large number is that we want to deal with systems ultimately
simulating our whole universe or continuous space-time manifolds (see
the more detailed discussion in \cite{lumps}). We first calculate
$r_{0}$. 
\begin{equation}
\begin{array}{|l|c|c|c|c|c|c|c|c|c|}
p & 0.9 & 0.8 & 0.7 & 0.6 & 0.5 & 0.4 & 0.3 & 0.2 & 0.1\\
\hline r_{0} & 4370 & 2063 & 1291 & 901 & 664 & 502 & 382 & 286 & 200
\end{array}\label{tabelle1}
\end{equation}
(for reasons we do not understand, we made some numerical errors in
the original table 1 in \cite{lumps}, p.2043. The correct numerical
calculations can be found in \cite{Requren}). It is more complicated
to give numerical estimates of the distribution of cliques, that is
$\langle Z_{r}\rangle$. After some manipulations and approximations
we arrived (in \cite{lumps}, p.2051f)
at the following approximate
formula and numerical table (the numerical values are given for $p=0.7$;
note that for this parameters the maximal order of occurring cliques,
$r_{0}$, was approximately $1291$) 
\begin{equation}
\log(\langle Z_{r}\rangle)\approx r\cdot\log(n)+n\cdot\log(1-p^{r})+r^{2}/2\cdot\log(p)
\end{equation}
(with $r^{2}/2$ an approximation of $r(r-1)/2)$ for $r$ sufficiently
large).\\[0.3cm]  
\begin{equation}
\begin{array}{|c|c|c|c|c|c|c|c|}
r & 600 & 650 & 800 & 1000 & 1200 & 1300 & 1400\\
\hline \log(\langle Z_{r}\rangle) & -5.7\cdot10^{6} & 3.2\cdot10^{4} & 3.2\cdot10^{4} & 2.5\cdot10^{4} & 8.4\cdot10^{3} & -0.75\cdot10^{2} & -1.1\cdot10^{4}
\end{array}\label{tabelle2}
\end{equation}
(In the original table 2 of \cite{lumps} the numerical values for
small and large $r$'s, lying outside the interval $(r_{0}/2,r_{0})$,
were wrong as we neglected numerical contributions which are only
vanishingly small in the above interval. The correct calculations
are taken from \cite{Requren}). The above table nicely illustrates
how fast the frequency of cliques of order $r$ drops to zero outside
the above interval.

As to the interpretation of these findings, one should remind the
reader that the above results apply to the generic situation, that
is, do hold for typical graphs (in very much the same sense as in
corresponding discussions in the foundations of statistical mechanics).
An evaluation of the combinatorial expressions in this and the following
sections shows that frequently the same kind of extreme probabilistic
concentration around, for example, \textit{most probable values} occurs
as in ordinary statistical mechanics. 

What is not entirely clear is, how far the random graph approach can
be applied to our complex dynamical networks.
Our working philosophy is that these results serve to show what we hope to be the qualitative behavior of such systems.
As our systems follow deterministic dynamical laws, starting from
certain initial conditions, the behavior cannot be entirely random
in the strict sense. This holds the more so since we expect the systems
to evolve towards \textit{attracting sets} in phase space and/or generate
some large scale patterns. On the other hand, due to the constant
reorientation of the edges, being incident to an arbitrary but fixed
vertex and the generically large vertex degrees of the vertices, one may assume that the system is sufficiently random on small scales so that the random graph picture reproduces at least the qualitative behavior of such extremely complex systems.

To make this picture more quantitative, the general strategy is the
following. We count the typical number of active edges in our evolving
network at a given clock time $t$, calculate from this the corresponding
edge probability, $p(t)$, and relate this snapshot of our network
to a random graph with the same! edge probability. This should yield
at least some qualitative clues. That is, we expect that qualitative
characteristics of our evolving network can, at each given clock time,
be related to the characteristics of a corresponding random graph.
In this specific sense, one may regard the \textit{edge probability},
$p(t)$, as the crucial dynamical parameter of our network, regarded
as a statistical system.
\section{A Geometric Renormalization Group and the Continuum Limit of Discrete Geometries}
\label{Ren}In the preceding section we introduced the notion of \textit{cliques}
as maximal complete subgraphs of a given (random) graph. We are interested
in them because physically they represent lumps of vertices which
are maximally entangled among each other. That means, viewing the
graph as the geometric substrate on which our dynamical network lives,
the vertices or the respective DoF of a clique are directly interacting
with each other. Invoking the Wilson/Kadanoff picture of the \textit{renormalization
group} in statistical mechanics these cliques are assumed to act as
dynamical entities of their own in the larger network.

In this context we want to mention an older approach developed by
Menger and coworkers some time ago and which we discussed in greater
detail in \cite{Roy}. Ideas in this direction (a geometry of \textit{lumps}
in which points are not primary entities) were also briefly mentioned
in Mengers contribution to the anthology: Albert Einstein:Philosopher
Scientist, volume II (\cite{Einstein}); note also the comments of
Einstein in the same volume. 
\begin{bem}
The cliques in our dynamical networks may change their shape under
the imposed dynamics which may create and/or delete edges. This was
one of the reasons why we developed the concept of \emph{fuzzy lumps}
or \emph{fuzzy cliques} in \cite{Roy}.
\end{bem}
Similar to the \textit{block spin} approach in the theory of critical
phenomena we promote the cliques of the initial (random) graph, $G$
(level zero), to the vertices of the
graph of the
next level (level one). We draw an edge between these vertices of
level one if the corresponding cliques have an overlap of a certain
order of vertices of level zero (see below). In this sense we get
a new graph, the so-called \textit{clique graph}, $Cl(G)$ or $G_{cl}$.
We can repeat this process, i.e. we can form the graphs $Cl(G),\, Cl^{2}(G)=Cl(Cl(G))\ldots$
and study the change of geometric characteristics and/or emergence
of new collective patterns on the various levels of this process.

So what is the physical picture underlying this process? The idea
of the ordinary (\textit{infrared}) renormalization process in e.g.
condensed matter physics is to integrate out the microscopic details
of the model and its dynamics and, via \textit{coarse graining} and
\textit{rescaling}, make visible the collective and large scale properties
of the model. This is accomplished by integrating out in each step
the DoF in the blocks or lumps of the preceding level and make the
new averaged DoF into the constituents of the next level, thus establishing
a new coarse grained and rescaled model together with a new coarse
grained Hamiltonian.

Slightly deviating from this philosophy, we want to concentrate in
our geometric renormalization process, for the time being, primarily
on the underlying geometric substratum. While we presume that one
can treat e.g. gravitation as being emergent from some dynamical network model
as we have introduced it in section \ref{network} (including the existence of a metric field $g_{ij}(x))$, we want to deal at the moment only
with the geometric large scale properties of our dynamic graphs.
\begin{bem}
\label{rem6.2}The following point is important. In the ordinary renormalization
process in for example some spin models, the spins in a certain block
are regarded as behaving in a coherent way, i.e., provided the \emph{correlation
length} is larger than the diameter of the blocks, the spins in some
block are expected to be more or less aligned. Therefore it makes
sense to build an average spin as a new block variable. In our network
we want, on the contrary,
to implement not only the wild \emph{vacuum fluctuations} which are expected to be very large on small scales but also create some smooth classical space-time continuum on large scales.

As a consequence of the dynamical laws we introduced above, the individual
vertex states in a certain clique fluctuate considerably but, due
to the close entanglement induced by the edges, we expect them to
\emph{cooperate} strongly so that some global coherent pattern may
emerge. The picture we have in mind is that of an array of coupled
\emph{phase oscillators} as will be described in the following section
about \emph{wormhole spaces} and \emph{small world networks}.\end{bem}
\begin{ob}
As a result of the process described in the following, we arrive at
a picture where lumps or cliques are contained in larger cliques (of
the next level) which are again contained in larger cliques of the
following level etc. So we get a hierarchical picture of the concept
of \emph{physical points} which have a rich internal structure.\end{ob}
\begin{defi}
We want to denote this double structure of an underlying erratic network
coexisting with a conjectured coherently behaving smooth classical
macroscopic surface structure by \emph{QX/ST} (\emph{quantum space}
plus classical space time). 
\end{defi}
We talked in
in subsection \ref{random}
about the existence of threshold functions which resemble phase transitions
and phase transition lines in statistical mechanics. Furthermore,
the different phases are frequently characterized by non-vanishing
\textit{order parameters}. In this spirit we make the following definition: 
\begin{defi}

If such a super structure ST emerges in our dynamical network in a
certain parameter regime we call it an \emph{order parameter manifold}.
\end{defi}
\begin{bem}
This implies that the existence of such a ST is a particular property
prevailing only in a certain region of the phase space of the model,
that is, we may also have such a microscopic substratum without an
overlying classical space time.
\end{bem}
This is, to some extent, similar to the situation in some other mainstream theories about the structure of spacetime like loop quantum gravity. There, a certain state of a spatial region (in the canonical formulation) is described by a certain vector in a Hilbert space. This vector in general corresponds to a certain graph (which is a superposition of basis states or graphs called spin networks) which is considered to underly that region. The state (or the corresponding graph) might or might not correspond to any “classical” space. There are very strong indications that for example singular regions like the ones inside a black hole correspond to states that do not have any classical counterpart. In other words those states are highly quantum and there is no classical notion of space that can be attached to them. Our model also seems to have the capability of incorporating this idea in a rather natural way.

It is interesting
to consider
what takes the place of the so called \textit{critical} systems of
the ordinary renormalization group. Critical systems converge to \textit{fixed
points} which, on their part, describe the large scale properties
of the critical system under discussion. 
\begin{conjecture}
In our geometric case, where cliques and their entanglements are expected
to be the crucial building blocks, geometric critical systems are
conjectured to display some geometric \emph{long range order} and
a certain \emph{selfsimilarity}. This will be worked out in more detail
in the following.
\end{conjecture}
It turns out that it is advisable to split the investigation into
two subsections. In the first one which deals with the clique structure,
we develop the geometric coarse graining process. In the second section
we discuss the rescaling process which leads to a \textit{continuum
limit}. Both parts of the renormalization process have problems of
their own and lead to quite deep mathematics. The material of the
following two subsections consists mainly of a review of the content
of the papers \cite{lumps},\cite{Requren}, parts of \cite{Requscale}
and \cite{Requcon}. That is due to length limitations we have to
refer the reader to these papers as to the partly quite complicated
and intricate combinatorial and numerical calculations. 
\subsection{The Geometric Coarse-Graining}
As indicated above, we assume that presently we live in a network
scenario where in the underlying microscopic network, called $QX$,
there exists a superstructure or \textit{order parameter manifold},
$ST$ (classical space-time). The emergence of $ST$ signals the transition
from a chaotic initial phase, $QX_{0}$, to a phase developing a \textit{near/far-order},
i.e., a \textit{causal structure} and relatively stable \textit{physical
points} or \textit{(fuzzy) lumps} (Menger). 

Our physical picture concerning the initial scenario (\textit{big
bang era}) is the following. The network $QX$ started from a presumably
densely entangled initial phase $QX_{0}$, in which on average every
pair of vertices, $x_{i},x_{k}$ is connected by an active edge with
high probability $p\approx1$ or a $J_{ik}\neq0$ (in our examples
of dynamical laws).

We then envisage two main epochs of our evolving network, a so-called
\textit{embryonic epoch} with a still large edge probability near
$p=1$ and an \textit{unfolded epoch} with a much smaller value of
$p$ and a large diameter
(see below for the definition)
at least on some higher \textit{clique level}. 
\begin{ob}
One should note that pure random graphs have a very small \emph{diameter}
for a large range of $p$-values (cf. e.g. \cite{Bollo2}) so that
$p$ must be quite small if we want to have an unfolded epoch. \end{ob}
\begin{defi}
The diameter of a graph is defined by $\max_{x_{i},x_{k}}\, d(x_{i},x_{k})$.
\end{defi}
In the subsection \ref{random} we calculated the typical order and
number of cliques and provided two tables for the order of the largest
expected cliques (the \textit{clique number} of the random graph),
$r_{0}$, and number of $r$-cliques. The numerical values were $p=0.7,\, n=10^{100}$.
We showed that almost all cliques have an order between $r_{0}$ and
$\frac{r_{0}}{2}$.

In a first step we want to clarify the mathematical and physical process
of constructing the clique graph $Cl(G)$ of a graph $G$. 
\begin{defi}
We employ two methods of constructing the clique graph:
\begin{enumerate}
\item The mathematical clique graph operator $Cl_{m}$ is defined as follows.
The vertices of the (mathematical) clique graph are the cliques of $G$, an edge is drawn if two cliques have a non-empty overlap.
\item The physical clique graph operator $Cl_{p}$ is constructed in the
following way. We delete too small (marginal) cliques which do not
lie in the above described interval $(r_{0},r_{0}/2)$. An edge between
cliques $C_{i},C_{k}$ is drawn if the overlap is non-marginal (we
discuss the physical implications of this notion below). That is,
we say, the overlap of the respective cliques is non-marginal if it
is larger than a certain value $l_{0}(r)$ which depends on $r$ and
the underlying physics.
\end{enumerate}
\end{defi}
\begin{ob}
In this way the original graph and the mathematical clique graph are
\emph{purified}, put differently, the iterated mathematical clique
graph is \emph{coarse-grained}, that is, on each level some marginal
structure is deleted. \end{ob}
\begin{bem}
To give examples for $l_{0}$; in \cite{Requren} section V.B we took
$n=10^{100},\, p=0.7$ which yielded $r_{0}=1291$. Our choice for
$l_{0}$ was $l_{0}=50$. We however convinced ourselves that the
physical picture does not critically depend on this choice.
As an example, for $l_{0}=30$ we got similar results.
\end{bem}
The physical motivation behind this procedure is the following. Our
aim is it to construct a space-time substratum in which we have, on
the one hand, a \textit{near order structure} concerning interaction
and the flow of information among neighboring physical points of $ST$
(i.e., the classical space-time concept under a certain magnification,
so that the internal structure of the points of the manifold becomes
visible). On the other hand, we have \textit{weak, translocal ties}
between lumps, which are with respect to the ordinary space-time metric
some distance apart. These weak ties result from edges which have
been present in the original graph $G$
or on some lower clique level, say $Cl^{m}(G)$, relative to some
clique level $Cl^{n}(G)$ with $n>m$ or a level which is already
near our presumed classical space-time $ST$ and have been neglected in the construction of the consecutive levels of the physical, purified clique graph. This dual picture will
become clearer in the next section on wormhole spaces and small world
networks. 
\begin{conjecture}
We conjecture that these weak translocal ties are responsible for
some important aspects of quantum theory as for example {\em entanglement}. 
\end{conjecture}
In this context we regard the cliques of some given level as approximately
autonomous subunits which are coupled both with their neighbors via
strong ties and with more distant cliques via some weak ties. We conjecture
that this dual structure will lead to a sort of global coherence which
results in the emergence of the order parameter manifold $ST$. 

We now want to briefly discuss some analytic and numerical results
of our construction of the (purified) clique graph. For the details
of the mostly quite intricate calculations we refer to the above mentioned
papers. We begin with the notion of the \textit{embryonic epoch} (cf.
subsection 4.1 of \cite{lumps}). In this epoch most of the edges
are still active, i.e. $p\approx1$. In subsection 4.1 of \cite{lumps}
we dealt with the question that under what conditions all the cliques
of the graph do have a common nonempty overlap. On the other hand,
for sufficiently small $p$ we could show that this overlap is empty
with high probability. This last regime describes presumably the so called \textit{unfolded epoch}.
The main work consists of providing \textit{combinatorial/probabilistic}
expressions for these graph properties.

In a next step, we want to calculate the order of the so called \textit{local
group} of a fixed clique. This local group comprises of the cliques
which have non-marginal overlap with an arbitrary but fixed clique,
$C_{0}$. This is an interesting graph property in the unfolded epoch
because it defines the infinitesimal
(first)
neighborhood of a vertex in the clique graph. The necessary combinatorial/probabilistic
analysis is made in subsection 4.1 of \cite{lumps} and section V.B
of \cite{Requren}. The number of cliques overlapping with some given
clique $C_{0}$ is, as all these properties, a random variable in
our random graph picture. We denote it by $N(C_{0};r',l)$ and its
expectation value by $<N(C_{0};r',l)>$. In this expression $C_{0}$
is a clique of fixed order $r$. Its overlap is considered with respect
to cliques of some order $r'$, both lying in the interval $(r_{0},r_{0}/2)$
i.e. $r_{0}<r,r^{\prime}<r_{0}/2$
. The overlap is denoted by $l\geq l_{0}$ with $l_{0}$ some minimal
value (denoting the non-marginal overlap). In the end this expression
is summed over all $r'$ lying in the interval $(r_{0},r_{0}/2)$.

In the same way we analyze the other defining parameters of the clique
graph (cf. section V.B of \cite{Requren}). 
\begin{bem}
One should note that one can infer from our second table in subsection
\ref{random} that, generically, there do exist much more cliques
than vertices in a typical random graph, i.e., $Cl(G)$ has typically
much more vertices than $G$. \end{bem}
\begin{ob}
Our numerical analysis shows after some intricate calculations and
estimates that the clique graph of a random graph $G$ with 
\begin{equation}
n=10^{100}\;,\; p=0.7\;,\;\text{clique overlap}\; l\geq l_{0}=50
\end{equation}
has typically $10^{10^{4}}$ vertices, the edge probability is approximately
$10^{-7\cdot10^{3}}$ and the vertex degree is of the order of $10^{0.3\cdot10^{4}}$.\end{ob}
\begin{bem}
Note that the small edge probability is compensated by the huge vertex
degree.
\end{bem}
Proceeding in the same way, while however readjusting the non-marginal
overlap $l_{0}(r)$ on each level, we get the \textit{iterated purified
clique graph}. We convinced ourselves that the gross parameters of
the clique graphs on the consecutive levels seem to reach stable values
after only a few renormalization steps. 
\begin{ob}
As the whole scenario is much more irregular in our case compared
to the situation in e.g. statistical mechanics, the existence of a
fixed limit network under the geometric renormalization group (more
properly, modulo graph isomorphisms) should presumably be refined
a little bit. We introduce the important concept of \emph{quasi-isometry}
or \emph{coarse isometry} in the following subsection. We hence expect
that the limit of an iterated clique graph is only invariant under
the physical clique operator $Cl_{p}$ modulo quasi-isometries.
\end{ob}
\begin{defi}
An ordinary \emph{graph isomorphism} $I$ is a bijective map between
graphs so that the graphs are form-invariant, i.e., an edge exists
between two image vertices iff there exists an edge between the pre-images. \end{defi}
\begin{bem}
~
\begin{enumerate}
\item In section VI of \cite{Requren} we gave simple examples which show
that our expectations are apparently not completely far fetched. These
examples should however not be regarded as examples of possible space-times.
We see that both limit points and limit cycles are possible.
\item In pure mathematics the study of the iterated clique graph is indeed
an interesting field in graph theory. cf. for example \cite{Pizana1}, \cite{Pizana2}, \cite{Pizana3}.
\end{enumerate}
\end{bem}
We want to close this subsection with giving some important results
for the unpurified mathematical clique graph. We begin with the graph
property \textit{connectedness}. 
\begin{lemma}
If $G$ is connected (i.e., every two vertices are connected by a
path), the same holds for $Cl_{m}(G)$. 
\end{lemma}
Proof: Let $C,C'$ be two cliques and $x,x'$ two vertices with $x\in C\,,\, x'\in C'$.
There exists a path connecting $x$ and $x'$. We denote the corresponding
vertices by $x=x_{0},x_{1},\ldots,x_{k}=x'$. There exist cliques,
$C_{i}$, containing the pairs $(x_{i},x_{i+1})$ with $i=0,\ldots,k-1$.
Note that the cliques $C_{i},C_{i+1}$ have non-zero overlap. This
proves the statement. 
\begin{bem}
The property of connectedness need of course no longer hold for the
coarse-grained clique graph, $Cl_{p}(G)$. In the random graph framework
this property becomes a random variable and its probability can be
calculated (cf. e.g. Conclusion 5.10 in \cite{Requren}).
\end{bem}
Another interesting property is the possible change of distance under
the clique operator. The following lemma shows that distances do not
change significantly. 
\begin{lemma}
Let $G$ be a connected graph and $C,C'$ two cliques in $Cl_{m}(G)$,
having the distance $d_{Cl(G)}(C,C')$. Then we have 
\begin{equation}
d_{Cl(G)}(C,C')=\min_{x,x'}\, d(x,x')+1\;,\; x\in C\,,\, x'\in C'
\end{equation}

\end{lemma}
Proof: The proof is essentially already contained in the proof of
the following theorem (cf. theorem 7.4 in \cite{Requren} or see \cite{Pizana3}).
Choose a minimal-length path in $Cl_{m}(G)$ connecting $C$ and $C'$
and consisting of the cliques 
\begin{equation}
C=C_{0},C_{1},\ldots,C_{k}=C'\quad\text{i.e.}\quad d_{Cl(G)}(C,C')=k
\end{equation}
implying that $C_{i},C_{i+1}$ have non-empty overlap. Choose a path
in $G$ starting at some $x_{1}\in C_{0}\cap C_{1},\,\text{with}\; x_{2}\in C_{1}\cap C_{2},\ldots$.
This is a path with initial point in $C$, endpoint in $C'$, having
the length $k-1$. It is easy to show that such a path is minimal
under this condition, i.e. it holds $k=(k-1)+1$, which proves the
statement.

The following observation is non-trivial and important. 
\begin{theo}
We assume that $G$, having a globally bounded vertex degree, has
scaling dimension $D$ (cf. section \ref{dim}). It follows that $Cl_{m}(G)$
has also a dimension with 
\begin{equation}
D_{Cl_{m}(G)}=D.
\end{equation}
\end{theo}
The proof is more intricate and can be found in section VII of \cite{Requren}. 
\begin{bem}
While there exists the restriction of a globally finite vertex degree, the
result shows that when building the iterated unpurified clique graph,
the microscopic dimension remains fixed in the transitions to higher
levels. In other words, if we surmise that our physical space-time
is smoother than the underlying microscopic substratum, implying among
other things that also the graph dimension on the different levels
of the iterated clique graph is expected to vary, we have to employ
the concept of the physical clique graph operator $Cl_{p}$ with its
purification and coarse-graining properties. That it is not easy to
change the graph dimension at all is shown in section VIII of \cite{Requren}.
We will briefly come back to this point in the following section.\end{bem}
\subsection{The Rescaling Process and the Continuum Limit}
In the preceding subsection we remained within the class of discrete
networks or graphs. I.e., at all levels of our constructions, the
models under discussion were discrete. We now will develop the framework
which allows to construct continuum limit models of our discrete networks.
While such a process is perhaps transparent in the context of models
living on a periodic Bravais lattice 
,it becomes very intricate in the case of general irregular networks
with a relatively deep amount of mathematics being involved. The general
context is the theory of general \textit{metric spaces}. We shall
make heavy use of material being developed for example in \cite{Gromov},\cite{Bridson}
and \cite{Harpe}. This subsection reviews
the content of \cite{Requcon} where more references can be found.

An important conceptual tool is the notion of \textit{quasi-isometry}.
This is the appropriate generalization of the notion of isometry to
disordered and irregular spaces where marginal details and variations
are partly ignored. 
\begin{defi}
Let $F$ be a map from a metric space, $X$, to a metric space, $Y$,
with metrics $d_{X},d_{Y}$ respectively. It is called a quasi-isometric
embedding if the following holds: There exist constants, $\lambda\geq1,\epsilon\geq0$,
such that
\begin{equation}
\lambda^{-1}\cdot d_{X}(x,y)-\epsilon\leq d_{Y}(F(x),F(y))\leq\lambda\cdot d_{X}(x,y)+\epsilon
\end{equation}
If, furthermore, there exists a constant $\epsilon'$ such that for
all $y\in Y$ we have $d_{Y}(y,F(X))\leq\epsilon'$, that is, $Y\subset U_{\epsilon'}(F(X))$,
(for the defintion of $U_{\epsilon}(A)$ see below) $F$ is called a quasi-isometry; the spaces are then called quasi-isometric.
There is an equivalent definition which shows that the preceding definition
is in fact symmetric between $X$ and $Y$ (see for example \cite{Harpe}).
That is, there exists a quasi-isometric map $G$ from $Y$ to $X$
with corresponding constants and $d_{X}(G\circ F(x),x)\leq\rho$ and
$d_{Y}(F\circ G(y),y)\leq\rho$ for some $\rho$. If $\lambda=1$
it is called a rough isometry.
\end{defi}
It is an important observation that in our framework of networks and
graphs many properties are stable under quasi-isometries. We have,
for example, the following results:
\begin{ob}
The ``growth type'' of graphs with globally bounded vertex degree
is stable under quasi-isometry (cf. section \ref{dim} and \cite{Harpe}).
We have in particular that quasi-isometric graphs have the same dimension,
i.e. (with $G_{1},G_{2}$ quasi-isometric graphs) 
\begin{equation}
\bar{D}_{1}=\bar{D}_{2}\quad,\quad\underline{D}_{1}=\underline{D}_{2}
\end{equation}
(see \cite{Requ2} for a proof).
\end{ob}
A further interesting observation is the following. We introduced
in section \ref{dim} the concept of local edge insertions/deletions
and showed that these procedures do not alter the dimension in the
case of graphs with globally bounded vertex degree. We have
\begin{ob}
Local edge insertions/deletions lead to quasi-isometries. By the same
token we see that via this method we get a rich class of examples
of quasi-isometric graphs.
\end{ob}
It is of great interest to derive criteria under which a network has
a finite growth degree (in contrast to e.g. an exponential growth)
and, a fortiori, an integer dimension. This is discussed in greater
detail in section 3 of \cite{Requcon}. We want furthermore to emphasize
that perhaps rather surprisingly our physically motivated interests
turn out to be closely related to a field of pure mathematics, i.e.
\textit{geometric group theory} via the concept of \textit{Cayley
graph} (cf. \cite{Harpe}). 

We now come to the construction of the continuum limit of an infinite
graph (for finite graphs the construction is not particularly interesting
as it leads to a single point). To this end we have in a first step
to construct a metric on a set of graphs or set of subsets of a certain space
(i.e., a metric on a set of metric spaces). We begin with the definition
of the so called \textit{Hausdorff-metric} on a space of subsets of
a metric space. 
\begin{defi}
Let $X$ be a metric space, $U_{\epsilon}(A)$ the $\epsilon$-neighborhood
of a subset $A\subset X$ ($U_{\epsilon}(A):=\{y:d(x,y)<\epsilon\,\text{for some}\; x\in A\}$.
The Hausdorff-distance between $A,B\subset X$ is then given by
\begin{equation}
d_{H}(A,B):=\inf\{\epsilon;A\subset U_{\epsilon}(B),B\subset U_{\epsilon}(A)\}.
\end{equation}

\end{defi}
We have the following lemma
\begin{lemma}
With $X$ a compact metric space, the closed subsets of $X$ form
a compact (i.e. complete) metric space with respect to $d_{H}$ (see
e.g. \cite{Bridson} or the book by Edgar;\cite{Edgar}).
\end{lemma}
In the following it is sometimes useful to make a slight generalization
to pseudo metric spaces as we will encounter situations where spaces
or sets have zero Gromov-Hausdorff-distance (for example, one being
a dense subset of the other) while they are not strictly the same.
Everything we will state for metric spaces in the following will also
hold for pseudo metric spaces. 
\begin{defi}
A pseudo metric fulfills the same axioms as a metric with the exception
that $d(a,b)=0\,\to\, a=b$ does not necessarily hold.
\end{defi}
The above distance concept is too narrow to be useful in a more general
context. It was considerably generalized by M.Gromov in an important
way (see \cite{Gromov1}) and later slightly modified by himself and
other authors (\cite{Gromov},\cite{Bridson},\cite{Petersen}). What
is really beautiful in our view is that while it seems to be more
abstract, it encodes the really important and crucial aspects of similarity
or ``nearness'' of spaces in a more satisfying way. That is, it
measures their structural similarity and not simply the nearness of
two structureless sets of points in a space. In general it is a pseudo
metric which may even take the value infinity. For compact spaces
it is always finite. If one forms equivalence classes of compact spaces
under isometries, it becomes a true metric.

The Gromov-Hausdorff distance, $d_{GH}$, can be formulated in two
equivalent ways. 
\begin{defi}
$d_{GH}(X,Y)$ between two metric spaces, $X,Y$, is defined as the
infimum of $d_{H}^{Z}(f(X),g(Y))$ over all metric spaces $Z$ and
isometric embeddings, $f,g$, of $X,Y$ into $Z$.\\ Equivalently,
one can define $d_{GH}$ by the infimum over $d_{H}(X,Y)$ in $X\sqcup Y$
(disjoint union) equipped with the metrics $d_{X\sqcup Y}$ which
extend the respective metrics $d_{X},d_{Y}$ in $X,Y$.
\end{defi}
The crucial part of the distance concept is always the triangle-inequality.
Furthermore we have to show that the above infimum is again a metric.
This is proved in section 4 of \cite{Requcon} where quite a few more
interesting results are discussed. Due to the lack of space we directly
embark on the deep results of Gromov concerning the formulation of
convergence of spaces towards each other. 

We now present the fundamental Gromov-compactness theorem, first for
compact spaces, then for more general cases.
\begin{defi}
We call a family of compact spaces, $X_{\lambda}$, uniformly compact
if their diameters are uniformly bounded and if for each $\epsilon>0$,
$X_{\lambda}$ is coverable by $N_{\epsilon}<\infty$ balls of radius
$\epsilon$ independent of the index $\lambda$. \end{defi}
\begin{theo}
(Gromov) A sequence $\{X_{i}\}$ contains a convergent subsequence
in $d_{GH}$ iff $\{X_{i}\}$ is uniformly compact.
\end{theo}
Proof: see \cite{Gromov1},\cite{Petersen} or \cite{Bridson}. Typically
an \textit{Arzela-Ascoli-Cantor-diagonal-sequence}-like argument is
used in the proof.

In our framework we are mainly interested in infinite graphs, i.e.
non-compact metric spaces being however frequently \textit{proper}. 
\begin{defi}
A metric space, $X$, is called proper if all its closed balls, $B(x,r)$,
are compact.
\end{defi}
We can then extend the above result in the following way. Ordinary
GH-convergence works well in the category of compact metric spaces.
If the spaces are non-compact, a slightly modified approach is more
satisfactory. One problem which may arise is that things in unbounded
spaces can ``wander away'' to infinity. So it is reasonable to pin
down the members of the sequence of spaces at certain points (base points), so that
they can be better compared. More precisely, we work in the category
of \textit{pointed} metric spaces, $(X,x)$, which is, a fortiori 
pretty normal from the physical point of view as it is like introducing
a reference point or a coordinate system.
\begin{defi}
The sequence of pointed metric spaces, $(X_{i},x_{i})$, is said to
converge to $(X,x)$ in pointed GH-sense if for every $r>0$ the sequence
of closed balls, $B(x_{i},r)$, converges to $B(x,r)$ in $d_{GH}$.
\end{defi}
The Gromov-uniform-compactness theorem now reads: 
\begin{theo}
If for all $r$ and $\epsilon>0$ the balls $B(x_{i},r)$ of a given
sequence $(X_{i},x_{i})$ are uniformly compact, then a subsequence
of spaces converge in pointed GH-sense.
\end{theo}
\begin{bem} There exist various slightly different notions of pointed
convergence in the literature. One can for example define pointed
GH-distance by admitting only isometries which map the base points
onto each other (\cite{Gromov}). Another possibility is to include
the distance of the images of the base points in the definition (\cite{Petersen2}).
The above definition is used in \cite{Bridson}. 
\end{bem}

We now apply these techniques to the following sequence of graphs.
We start with a graph, $G$, of globally bounded vertex degree, $v$,
and take $G$ with the original graph metric, $d$, as the initial
metric space. Then we generate a sequence, or more generally a directed
system of metric spaces, $\lambda G$, by taking the same graph, $G$,
but now with the scaled metric, $\lambda d$, defined as
\begin{equation}
\lambda d(x,y):=\lambda\cdot d(x,y)
\end{equation}
and (usually) taking $\lambda\to0$. One may, in particular, take
subsequences of the kind
\begin{equation}
G_{n}\;,\; d_{n}:=n^{-1}\cdot d\;,\; n\to\infty
\end{equation}
or replace $n$ by $2^{-k}$.

In a next step we have to show that all the above criteria are fulfilled
in this case which is a non-trivial task (see section 5 of \cite{Requcon}).
Among other things several new notions and concepts have to be introduced
like e.g. \textit{doubling measures} etc. Finally we can show that
our sequence of rescaled graphs has a continuum limit! It is now very
important to learn something about the structure of this limit space.
Some steps are done in section 5 of \cite{Requcon}. It is of particular
importance to understand under what conditions this limit space is
a smooth manifold or, on the other hand, a chaotic space of rather
fractal type. We are very interested in the possibility of a limit
space having a superficially smooth structure together with an internal
infinitesimal more erratic structure ``around'' the ``classical''
points of the base manifold, being kind of a generalized fiber space.
\section{Wormhole Spaces, Holography and the Translocal Structure of Quantum Theory}\label{wormhole-sp}
In this last short section, which is also kind of a conclusion, we want to discuss in a very sketchy way various fundamental (open) questions in modern physics and show how our above framework can be applied to them. As we shall treat these important topics in a quite cursory way we refer the interested reader to the following papers \cite{Requworm2},\cite{Requscale},\cite{Quantum} for a more thorough discussion.

In section \ref{Ren} we came already to the conclusion that presumably the cellular network substratum which has the propensity to lead to a continuum limit, resembling our physical (quantum) space-time, has to be in a peculiar critical state which resembles a \tit{scale free small world network} as we described it in \cite{Requscale}. This means that apart from a certain nearorder leading to a lumpy local structure in the network on the various scales of coarse graining and rescaling, there remains on all scales a certain \tit{sparse network} of the so-called \tit{translocal connections} (i.e., edges in the graph language relating distant regions with respect to the ordinary local metric).
Rather surprisingly such a structure was analyzed roughly at the same time in a quite different area of science, dubbed \tit{small world networks} (for a detailed treatment see e.g. \cite{Watts}. A brief discussion can also be found in \cite{Requloch}). 

In \cite{Requloch} we described such models in the following way.    
We start with a regular graph having a nearorder structure like e.g. $\Z^n$ with edges to nearest neighbors in the vertical and horizontal direction. In a scaling limit this would lead to a space like $\R^n$ with the Euclidean distance metric. In a next step we superimpose $\Z^n$ or a similar regular space with a sparse random graph on the same vertex set and a small edge probability $p$. These random edges now lead to an additional translocal structure on $\Z^n$. We conjecture that this serves as a toy model for the kind of substratum underlying our physical (quantum) space-time which we dubbed QX/ST (i.e. an underlying discrete substratum having both a near and a far order together with a continuous surface structure on a low resolution of space-time).
\begin{ob}A characteristic of small world networks is their surprisingly small diameter or mean distance (cf. e.g. \cite{Requscale},\cite{Requloch} or \cite{Watts}), given the sparseness of  additional translocal edges.
\end{ob} 
In \cite{Requworm2} we developed and studied this phenomenon in quite some detail and coined the notion \tit{wormhole spaces} for such structures like our QX/ST. We showed that the \tit{BH entropy area law} and the \tit{holographic principle} follow quite naturally from our framework.  

Finally, in \cite{Quantum} we undertook to develop a framework which describes the quantum phenomena as emergent properties on a mesoscopic scale. More precisely, at various places in our paper we indicated that our dynamic network of local lumps or cliques can be associated to a network of coupled \tit{phase oscillators} (cf. remark\ref{rem2.6}
or remark \ref{rem6.2}).
\begin{bem}Many non-linear systems approach limit cycles in their evolution on which they then evolve according to a law like $\dot{\theta}=\omega$, $\omega$ a certain specific {\em natural} frequency, $\theta$ the phase of the system. As to more details see \cite{Watts},\cite{Strogatz} or \cite{Quantum}.
\end{bem}  

The famous \tit{Kuramoto model} desribes a large population of coupled \tit{limit cycle} or \tit{phase oscillators} whose natural frequencies are drawn from some prescribed distribution (see for example \cite{Strogatz2}). The hallmark of such models is that there may occur a particular kind of phase transition in which all the initially different natural frequencies, $\omega_i$, become dynamically synchronized.
\begin{conjecture}We expect that our dynamical network models show a similar behavior with the cliques or lumps representing the limit cycle or phase oscillators.
\end{conjecture}

It is an important observation (\cite{Niebur}) that this emergent property of synchronization is strongly enhanced and stabilized by a certain sparse non-local network of random couplings superimposed on the prevailing network of local couplings in the array of oscillators. This is exactly what we found in our network models.
\begin{conjecture}This possibility of {\em phase locking} may be a hint how a global {\em time function} emerges from the array of initially different local times with the phase oscillators viewed as local clocks.
\end{conjecture} 

It is remarkable that Bohm, starting from a different direction, also speculates about the existence of a hierarchy of coupled oscillators on consecutive scales of (quantum) space-time (see \cite{Bohm}). He came to a conclusions similar to the ones uttered in \cite{Quantum} concerning the consequences for quantum theory as being emergent from such a deep structure. We discussed this in greater detail in \cite{Quantum}. 

\section{The (Quantum) Graphity Approach}\label{qgph}

In this section, we provide a brief review of another bottom-up approach
to quantum gravity, called quantum graphity \citep{Konopka1,Konopka2,Konopka3}. We divide the discussion into kinematics and dynamics.
In the next section, we try to give a brief comparison of the two approaches. 

\subsection{Kinematics}

\subsubsection{States and the Hilbert space}

Quantum graphity like some other bottom-up approaches starts from
a family of graphs living on a fixed set of vertices as underlying the notion of spacetime. This family comprises the complete graph $K_{N}$ (i.e., all vertices are linked to each other) and all the other possible graphs on the $N$ vertices. The vertices are distinguishable but are undirected, so for the edge $e_{ij}$ between two
vertices $x_{i}$ and $x_{j}$, we have $e_{ij}=e_{ji}$. A Hilbert
space $\mathscr{H}_{\textrm{edge}}$ is associated to each edge such
that 
\begin{equation}
\mathscr{H}_{\textrm{edge}}=\textrm{span}\left\{ |0\rangle,|1\rangle\right\} 
\end{equation}
where an edge being in a state $|0\rangle$ is considered ``off''
and one with a state $|1\rangle$ is considered an ``on'' edge.
This Hilbert space is the same as one that belongs to a fermionic
oscillator with $|0\rangle$ and $|1\rangle$ corresponding to no
particles or 1-particle states respectively. Another assumption is
that the degrees of freedom are only associated to edges and vertices
do not contribute to them. This means that the total Hilbert space
of a typical graph is the tensor product of the edge Hilbert spaces
$\mathscr{H}_{\textrm{edge}}$ such that 
\begin{equation}
\mathscr{H}_{\textrm{total}}=\bigotimes^{N(N-1)/2}\mathscr{H}_{\textrm{edge}}
\end{equation}
where $\frac{N(N-1)}{2}$ is the total number of edges in a complete
graph $K_{N}$ with $N$ vertices. As mentioned above, this total
Hilbert space is a space with members being the complete graph $K_{N}$
with $N$ vertices, and all of its subgraphs. Thus one can decompose
$\mathscr{H}_{\textrm{total}}$ as a tensor sum 
\begin{equation}
\mathscr{H}_{\textrm{total}}=\bigoplus_{G}\mathscr{H}_{G}
\end{equation}
over all subgraphs $G$ of the complete graph $K_{N}$, including
$K_{N}$.

\subsubsection{Operator representation}

In the next step, one defines the creation and annihilation operators
$a^{\dagger}$ and $a$ acting on the Hilbert space of each edge $\mathscr{H}_{\textrm{edge}}$
(and by extension, on $\mathscr{H}_{\textrm{total}}$) such that they
anticommute 
\begin{equation}
\{a,a^{\dagger}\}=1\label{eq:anti-commut-a}
\end{equation}
and 
\begin{equation}
a|0\rangle=0,\,\,\,\,\,\,\, a|1\rangle=|0\rangle,\,\,\,\,\,\,\, a^{\dagger}|0\rangle=1.
\end{equation}
One can also define the ``number operator'' for the edge $e_{ij}$
between the vertices $x_{i}$ and $x_{j}$ 
\begin{equation}
N_{ij}=a_{ij}^{\dagger}a_{ij}
\end{equation}
such that $N_{ij}$ acting on its eigenstates, i.e. ``number states''
that are the basis of $\mathscr{H}_{\textrm{edge}}$, yields 
\begin{equation}
N_{ij}|n_{ij}\rangle=n_{ij}|n_{ij}\rangle,\,\,\,\,\,\,\,\,\,\, n_{ij}=0,1
\end{equation}
which means that it returns an eigenvalue $0$ if the edge is off
and $1$ if it is on. Also note that since the graphs considered are
undirected, it follows that 
\begin{align}
a_{ij}= & a_{ji}\\
a_{ij}^{\dagger}= & a_{ji}^{\dagger}\\
N_{ij}= & N_{ji}.\label{eq:N-symm}
\end{align}
Then the authors make an observation that the matrix corresponding
to the eigenvalue of the number operator $N_{ij}$ is analogous to
the adjacency matrix $A_{ij}$ of an undirected graph in which an
element $A_{kl}$ is $1$ if there is an edge between the vertices
$x_{k}$ and $x_{l}$ and $0$ otherwise.

On the other hand, the powers of the adjacency matrix such as 
\begin{equation}
A_{i_{1}i_{n+1}}^{n}=\sum_{i_{1}}\cdots\sum_{i_{n}}A_{i_{1}i_{2}}A_{i_{2}i_{3}}\ldots A_{i_{n}i_{n+1}}
\end{equation}
contain information about the open (if $i_{1}\neq i_{n+1}$) and closed
(if $i_{1}=i_{n+1}$) paths between the vertices $x_{i_{1}}$ and
$x_{i_{n+1}}$ in the graph under discussion. So for example $A_{i_{1}i_{n+1}}^{n}$
denotes the number of possible paths between the vertices $x_{i_{1}}$
and $x_{i_{n+1}}$ by $n$ jumps or steps. But this does not necessarily
mean a non-overlapping path (or non-retracing path) of non-overlapping
length $n$, since one can jump back and forth over some of the edges
more than once. As an example they mention an element of the fourth
power $A^{4}$ of the adjacency matrix of a certain graph such that
$i_{1}=1,\, i_{2}=2,\, i_{3}=i_{1}=1,\, i_{4}=i_{2}=2,\, i_{5}=3$
and thus the path is between $x_{i_{1}}=x_{1}$ to $x_{i_{5}}=x_{3}$
and is represented by the sequence of edges $\{e_{12},e_{21},e_{12},e_{23}\}$.
This is an overlapping path with $4$ jumps or steps but with a non-overlapping
length of $2$, i.e. the minimum number of edges connecting $x_{i_{1}}=x_{1}$
to $x_{i_{5}}=x_{3}$ is only $2$ represented by $\{e_{12},e_{23}\}$.

In order to avoid this and only count the number of non-retracing
paths between two vertices, they make the following observation. In
a first step, one uses the number operator $N_{ij}$ instead of $A_{ij}$
to extract the information about the number of paths between two vertices.
Then, like in quantum field theory, a normal ordering, using the sign
$:N_{ij}^{n}:$ is used in the powers of $N_{ij}$. As usual it is
such that all the annihilation operators are brought to the right
of creation operators. For example for the second power we have 
\begin{equation}
N_{ik}^{2}=:N_{ij}N_{jk}:=:a_{ij}^{\dagger}a_{ij}a_{jk}^{\dagger}a_{jk}:=a_{ij}^{\dagger}a_{jk}^{\dagger}a_{ij}a_{jk}.
\end{equation}
Thus in the product appearing in $N^{n},\, n\geq2$, if any of the
edges appear more than once in the path, its number operator $N_{kl}$
will also appears more than once in the form $N_{kl}N_{lk}$ (which
is equal to $N_{kl}N_{kl}$ due to symmetry of $N$ in (\ref{eq:N-symm}))
in the product, and then we will have at least two of the (same) corresponding
annihilation and creation operators in the product as 
\begin{equation}
N^{n}=\cdots a_{kl}^{\dagger}a_{lk}^{\dagger}\cdots a_{kl}a_{lk}\cdots=\cdots a_{kl}^{\dagger}a_{kl}^{\dagger}\cdots a_{kl}a_{kl}\cdots=0
\end{equation}
and it will vanish due to the anticommutation (\ref{eq:anti-commut-a}).
This way the eigenvalues of the operator $N_{ik}^{n}$ return the
number of non-overlapping paths between the vertices $x_{i}$ and
$x_{k}$.

\subsection{Dynamics}

In bottom-up models, one starts from a discrete underlying structure
like a graph, and aims to derive structures like spacetime as an emergent
phenomenon through basic evolution laws on the discrete system, i.e.
on the graph. Then the evolution laws, or the Hamiltonian in the case
of quantum graphity, should be defined to make this possible in one
way or another. The guidelines for defining the evolution laws or
the Hamiltonian are generally based on educated guesses and having
the final result of evolution, i.e. emergence of some structure, in
mind. In quantum graphity, there are several guidelines to define
a Hamiltonian, including the followings:
\begin{enumerate}
\item There are two types of Hamiltonians: the first type are the ones that
just measure the amount of ``energy'' of a graph and do not change
the linking structure and topology of it, and are called the free
Hamiltonians. The second type are those that do change the linking
structure and topology of the graph and are called the interaction
Hamiltonians. The free Hamiltonians preserve the graph automorphism
symmetry, i.e. permutation symmetry: a map of graph $G$ onto itself,
$\sigma:G\rightarrow G$, such that the edge $(x_{i},x_{j})$ is in
the domain of $\sigma$, iff the edge $(\sigma(x_{i}),\sigma(x_{j}))$
is in the image of $\sigma$.
\item All the Hamiltonians are written in terms of the adjacency matrix
or equivalently the number operator and its powers. They also may
include the creation and annihilation operators. 
\item The evolution should have a preference for global vertex degree $\mathfrak{d}$
such that after a long ``time'', it settles to a state that is of
this nature.
\end{enumerate}
The full Hamiltonian of the system then would be the sum of all the
free and interacting Hamiltonians. In what follows, we describe in
some details the explicit form of the Hamiltonians, suggested by the
authors.

\subsubsection{The free Hamiltonians}

In their model, the authors introduce two free Hamiltonians, one related
to the valence or degree of vertices (although vertices themselves
do not carry any degree of freedom) and another one related to the
number of closed non-overlapping paths. 

For the first term associated to the vertex degree, there are many
possibilities that conform to the criteria 1 and 2 mentioned above,
such as the trace of the number operator, or the sum of the off-diagonal
elements of it. For item 3 above, first it is desirable to find an
expression, based on the number operator (or adjacency matrix) that
counts the number of edges attached to a vertex $x_{i}$, i.e. its
vertex degree. This expression is
\begin{equation}
v_{i}=\deg(x_{i})=\sum_{j}N_{ij}.
\end{equation}
It basically sums over the the elements of the $i$'th row in the
matrix of $N_{ij}$, i.e. counts the number of $1$'s in row $i$.
Using this, they seek to find a Hamiltonian in the form
\begin{equation}
H_{V}^{\prime}=g_{V}\sum_{i}f_{i}\left(\sum_{j}N_{ij},\mathfrak{d}\right)
\end{equation}
where $g_{V}$ is a positive coupling constant, and to make the ``energy''
of a graph with all the vertex degrees equal to $\mathfrak{d}$ a
minimum state (or in other words, the ``equilibrium point'' of the
evolution), the function $f_{i}$ is defined such that its minimum
occurs when vertex $x_{i}$ has a vertex degree $v_{i}=\mathfrak{d}$.
This way if all the vertices of the graph have a vertex degree exactly
equal to $\mathfrak{d}$, the function $H_{V}^{\prime}$ will have
its minimum. Based on these observations, the authors then introduce
a specific form of this class of Hamiltonians
\begin{equation}
H_{V}=g_{V}\sum_{i}e^{p\left(\mathfrak{d}-\sum_{j}N_{ij}\right)^{2}}
\end{equation}
where $p$ is a real constant. The exponential is defined by its series
expansion in $p$ and they claim that for each deviation of a vertex
degree from $\mathfrak{d}$ such as $v_{i}\neq\mathfrak{d}$, there
is a ``penalty'' in energy, i.e. a deviation from the minimum energy
which is roughly of the form
\begin{equation}
\delta E_{V}\sim e^{p\left(v_{i}-\mathfrak{d}\right)^{2}}.
\end{equation}
The second free Hamiltonians, $H_{B}$, measures the contributions
to the energy associated to a graph due to closed non-overlapping
paths in it. It is a sum over all vertices
\begin{equation}
H_{B}=\sum_{i}H_{B_{i}}
\end{equation}
such that the term $H_{B_{i}}$ is centered on vertex $x_{i}$ and
written in the form
\begin{equation}
H_{B_{i}}=-g_{B}\sum_{j}\delta_{ij}e^{rN_{ij}}=-g_{B}\sum_{j}\delta_{ij}\sum_{L=0}^{\infty}\frac{r^{L}}{L!}N_{ij}^{L}.\label{eq:HBi}
\end{equation}
Here $g_{B}$ is a another positive coupling constant, $r\in\mathbb{R}$
and $L$ is the length of a non-overlapping path. The term $N_{ij}^{L}$
returns the number of paths of length $L$ that connect vertices $x_{i}$
and $x_{j}$. Note that as mentioned before, $N_{ij}^{L}$ does not
count overlapping paths and also because of presence of the $\delta_{ij}$,
only closed path are counted that start and end on the same vertex.
The energy it assigns to a graph state $|\psi_{G}\rangle$
\begin{equation}
\langle\psi_{G}|:H_{B}:|\psi_{G}\rangle=E_{B}(G)
\end{equation}
is
\begin{equation}
E_{B}(G)=-\sum_{i}\sum_{L=0}^{\infty}g_{B}(L)P(x_{i},L)\label{eq:EB}
\end{equation}
where $P(i,L)$ is the number of closed paths of length $L$ at each
vertex $x_{i}$ and $g_{B}(L)$ is an effective coupling of the form
\begin{equation}
g_{B}(L)=\frac{r^{L}}{L!}g_{B}.
\end{equation}
Note that $L=1$ and $L=2$ terms in (\ref{eq:HBi}) will vanish since
there is no closed path with length $1$, and also any closed path
of length $2$ is an overlapping one which is prohibited by the properties
of $N_{ij}$ due to (\ref{eq:anti-commut-a}) as discussed before.
Terms with $L=0$ are also simple constants. So the interesting contributions
to $H_{B}$ start from $L=3$. Looking at (\ref{eq:EB}) we notice
that the more closed paths a graph has, the lower energy it possesses.
But since the graph is finite, the energy of the system is bounded
from below. Also due to the observation that the effective coupling
$g_{B}(L)$ in many cases falls faster than $P(x_{i},L)$, very large
closed paths do not contribute significantly to the energy of the
system. So $H_{B_{i}}$ is a ``quasilocal'' operator and also the
energy per vertex remains finite.

The system exhibits two scales $L^{*}$ and $L_{i}^{**}$ where $L^{*}$
is the number of closed paths that maximize the effective coupling,
i.e.
\begin{equation}
g_{B}\left(L^{*}\right)>g_{B}\left(L\right),\,\,\,\,\,\,\,\,\,\forall L\neq L^{*}
\end{equation}
and $L_{i}^{**}$ defined for each vertex $x_{i}$ is the number that
maximizes $g_{B}(L)P(x_{i},L)$, i.e.
\begin{equation}
g_{B}\left(L^{**}\right)P\left(x_{i},L^{**}\right)>g_{B}(L)P(x_{i},L),\,\,\,\,\,\,\,\,\,\forall L\neq L^{**}
\end{equation}
which is a graph dependent scale.

So this free Hamiltonian operator has information about the energy
of the system due to the number of closed paths with the minimum energy
favored by the scale $L^{**}$.

\subsubsection{Interaction Hamiltonian}

The interaction Hamiltonians are the ones that provide the evolution
of the states of the graph, and this evolution is what leads to the
emergence of specific structures or lead to stable states that can
mimic spacetime. One physical guideline for the authors of the model
is the locality of these interactions, meaning that given an edge
or a vertex, only a small neighborhood of them can be affected by
the evolution. As an example, the famous Game of Life has a local
evolution where only cells in the first neighborhood of a certain
cell are affected by the state of that cell. In the case of quantum
graphity, the following paragraphs will make this locality notion
more clear.

The authors divide the interaction Hamiltonians into two cases: $H_{(\textrm{add})}$
which describes the addition or subtraction of edges between three
vertices in which the vertex degree of at least one of the involved
vertices changes, and $H_{(\textrm{exch})}$ which describes the exchange
of already existing edges between four neighboring vertices, such
that the vertex degrees of the involved vertices remain invariant.

The exchange Hamiltonian is defined as
\begin{equation}
H_{(\textrm{exch})}=g_{(\textrm{exch})}\sum_{ijkl}{}^{\prime}N_{ij}\left(a_{il}^{\dagger}a_{jk}^{\dagger}a_{jl}a_{ik}\right)
\end{equation}
which as can be seen, already possesses the normal ordering mentioned
before. Here the prime on the summation indicates that the vertices
$x_{i}$ to $x_{l}$ involved in the Hamiltonian are all different.
It is seen easily that the evolution only involves four vertices that
are in a very close neighborhood of each other. This evolution deletes
the edges between the vertices $x_{i}$ and $x_{k}$ and between $x_{j}$
and $x_{l}$, and then adds two edges, one between $x_{j}$ and $x_{k}$
and one between $x_{i}$ and $x_{l}$, but due to the presence of
$N_{ij}$, all that will happen only if the link between $x_{i}$
and $x_{j}$ is already on.

The addition/subtraction Hamiltonian is introduced in the following
form
\begin{equation}
H_{(\textrm{add})}=g_{(\textrm{add})}\sum_{ijk}{}^{\prime}N_{ij}N_{ik}\left(a_{jk}+a_{jk}^{\dagger}\right).
\end{equation}
Here again, due to the presence of $N_{ij}$ and $N_{ik}$, given
that there is already an on link between $x_{i}$ and $x_{j}$, and
between $x_{i}$ and $x_{k}$, then a link between $x_{j}$ and $x_{k}$
is created if there was no link between them before, or it will be
deleted if it existed before.

In the following subsection, we discuss an extended version of the
model, similar to the older version of quantum graphity, where edges
have internal degrees of freedom and will review the idea that this
can be used to incorporate matter degrees of freedom.

\subsubsection{Inclusion of more degrees of freedom; matter}

The internal states of edges can be extended by letting them have
several ``on'' states $|1_{s}\rangle$ for $s\in\mathbb{N}$ instead
of just one, i.e $|1\rangle$. Of course the off state is still only
$|0\rangle$. Then the creation and annihilation operators acting
on the Hilbert space of each edge will also have an index $s$ such
that
\begin{equation}
\left\{ a_{s},a_{s^{\prime}}^{\dagger}\right\} =\delta_{ss^{\prime}}.
\end{equation}
This is similar to the case of several types of fermions in a fermionic
system. One can then define several Hamiltonians corresponding to
these new degrees of freedom. 

If one chooses $s\in\{1,2,3\}$, then the system can be cast into
a form such that it resembles a spin-1 particle and the states can
be represented by
\begin{equation}
\mathscr{H}_{G}=\textrm{span}\left\{ |j,m\rangle:j\in\{0,1\},\,-j\leq m\leq j\right\} .
\end{equation}
Then one can define operators acting on the Hilbert space of each
edge, similar to the $J_{z}$ and the ladder operators $J_{+}$ and
$J_{-}$ of the spin-1 particle.

Note that now the edges have internal states and these can be used
to describe matter degrees of freedom. This is done by assiciating
$m=0$ to ground states and $m=\pm1$ to the excited states interpreted
as matter.

\section{A Brief Comparison of the Two Approaches}\label{compare}

In this last section we undertake to describe the characteristics of the two frameworks. It serves at the same time as a conclusion.
\begin{itemize}
\item Quantum graphity (QGph) starts with a Hilbert space description and (anti)commutator
operator representation on it, which implies that quantum theory
(in a general sense; a description based on a Hilbert space etc.)
is assumed to be valid at the bottom-most layer of the universe, i.e. the universe
is fundamentally quantum. Furthermore the existence of a Hamiltonian is assumed and most of the calculations are performed by employing something like the canonical ensemble in a given heat-bath. In SDCN, on the other had, the description on the most fundamental level is instead inspired by the discrete dynamics of large disordered networks and cellular automata (CA).The quantum-like behavior of physics is expected to \emph{emerge} on a higher level. This holds also for most of the other continuum concepts.   
\item In QGph, the degrees of freedom are all associated to edges, and vertices
do not play an important role, while in the SDCN the dynamics and
the degrees of freedom correspond to both edges and vertices. Exclusion
of vertices from having degrees of freedom looks a bit unnatural.
It also limits the richness of the model and its ability to produce
more interesting physics and complex behavior. It is important that in SDCN the elementary vertex and edge states coevolve under the dynamics, following the philosophy of general relativity, that is, matter degrees of freedom (DoF) interacting with the geometric DoF. 
\item An important difference between the two models is the existence of
several coarse-grained layers in the SDCN compared to only one layer
in the QGph. In other words, in the SDCN, we not only have an evolution
in ``time'', but also at each step, there exist several layers in
the network (graph) that represent the effective physics at different scales of resolution
of the universe. In our view, this makes more sense due to the fact
that we already know that at different scales, physical descriptions
and variables used may be different. A simple example is the description
and the relevant variables in thermodynamics of a gas compared to
it statistical mechanics where the variables of the former can be
seen as coarse-grained variables of the latter and the dynamical laws
also look different.
\item The cornerstone of the SDCN-framework is the formulation of a geometric renormalization group, with the various steps consisting of coarse graining and rescaling. In this context advanced concepts in graph theory are employed like cliques and clique graphs. We argue that classical space-time and its microstructure (the vacuum fluctuations) occur as a fixed point of such a renormalization process.
\item We provide arguments that only particular network states can lead to such a classical limit. We call them critical network states. They are characterized by both a local nearorder structure plus a superimposed sparse translocal farorder structure. Ths picture resembles the famous small world networks. It leads to a finestructure of classical space-time which we dubb wormhole spaces. In contrast to this picture QGph assumes that the deep structure of space-time is built by quite regular latticelike graphs. This view is perhaps inspired by the spinnetwork picture in which only relatively simple graphs occur.
\item The existence of several layers in SDCN also allows us to explain, some bizarre properties of the quantum
realm such as superpositions, entanglements and the ``spooky action-at-a-distance''
as referred to by by Einstein. This follows from an analysis of the \tit{critical states} on the primordial network level which are capable to lead to a macroscopic (quantum) space-time on the surface level with a built-in translocal entanglement structure. Furthermore the emergence of holography and the BH entropy-area law is explained with the help of this translocal wormhole structure.
In QGph, since the fundamental description
is quantum, these issues are taken for granted or as-is and there
is no deeper explanation as to why these phenomena appear in the universe. 
\item Another issue is the notion of dimension. In the SDCN we have an explicit definition
of the dimension which can in general adopt arbitrary real values. It can remain invariant or vary after each coarse-graining step (in e.g. the critical network states). We also explore the condition that this dimension is for
example integer (these properties are connected with deep mathematical problems). In QGph there is no discussion of this kind and in
fact the notion is not explored.
\item Although the dynamics of both models are guided by educated guesses
and desired end results, it seems that the dynamics in QGph is more
fine tuned in the sense that the criteria for the equilibrium state
look more hand picked. The criteria for a ``minimum energy state
'' there, corresponds to a graph with certain vertex degree and number
of closed paths. The final or equilibrium state in the SDCN happen
in a more natural and less fine tuned way and it is related to the
more or less fixed patterns that appear at the highest coarse-graining
level (although they might happen in the lower levels too).
\end{itemize}
In the face of these observations we think that the two approaches are markedly different but may complement each other.


\begin{acknowledgments}
S.R. would like to acknowledge the support of the PROMEP postdoctoral fellowship (through UAM-I), the grant from Sistema Nacional de Investigadores of the CONACyT and the partial support of CONACyT grant: Implicaciones F\'{i}sicas de la Estructura del Espaciotiempo.
\end{acknowledgments}



\begin{thebibliography}{99}
\small{
\bibitem{Carlip}S.Carlip:``Challenges for Emergent Gravity'', arXiv:1207.2504
\bibitem{Konopka1}T.Konopka,F.Markopoulou,L.Smolin:``Quantum Graphity'', arXiv:hep-th/0611197
\bibitem{Konopka2}T.Konopka,F.Markopoulou,S.Severini:``Quantum Graphity: a model of emergent locality'', PR D77(2008)104029, arXiv:0801.0861
\bibitem{Konopka3}T.Konopka:``Statistical Mechanics of Graphity Models'', PR D78(2008)044032, arXiv:0805.2283
\bibitem{Hooft1}G.'t Hooft:``The Cellular Automaton Interpretation of Quantum Mechanics'', arXiv:1405.1548
\bibitem{Hooft2}G.'t Hooft:``Equivalence Relations Between Deterministic and Quantum Mechanical Systems'', J.Stat.Phys. 53(1988)323
\bibitem{Hooft3}G.'t Hooft:``Quantization of Discrete Deterministic Theories by Hilbert Space Extension'', Nucl.Phys. B342(1990)471
\bibitem{Wheeler1}C.W.Misner,K.S.Thorne,J.A.Wheeler:``Gravitation'', Freeman, San Francisco 1970
\bibitem{Requ1}M.Requardt:``Cellular Networks as Models for Planck Scale Physics'', J.Phys. A: Math.Gen. 31(1998)7997, arXiv:hep-th/9806135
\bibitem{Requ2}T.Nowotny,M.Requardt:``Pregeometric Concepts on Graphs and Cellular Networks'', (invited paper) Chaos, Solitons and Fractals 10(1999)469, arXiv:hep-th/9801199
\bibitem{Requ3}T.Nowotny,M.Requardt:``Dimension Thery on Graphs and Networks'', J.Phys. A: Math.Gen. 31(1998)2477, arXiv:hep-th/9707082
\bibitem{Requren}M.Requardt:``A Geometric Renormalization Group and Fixed-Point Behavior in Discrete Quantum Space-Time'', J.Math.Phys. 44(2003)5588, arXiv:gr-qc/0110077
\bibitem{Requcon}M.Requardt:``The Continuum Limit of Discrete Geometries'', Int.J.Geom.Meth.Mod.Phys. 3(2006)285, arXiv:math.ph./0507017
\bibitem{Requscale}M.Requardt:``Scale Free Small-World Networks and then Structure of Quantum Space-Time'', arXiv:gr-qc/0308089
\bibitem{Requworm}M.Requardt:``Wormhole Spaces, Connes' Points, Speaking to Each Other, and the Translocal Structure of Quantum Theory'', arXiv:hep-th/00205168
\bibitem{Requworm2}M.Requardt:``Wormhole Spaces: the Common Cause for the Black Hole Entropy-Area Law, the Holographic Principle and Quantum Entanglement, arXiv:0910.4017''
\bibitem{Ila1}A.Ilachinski:``Cellular Automata, a Discrete Universe'', World Scientific Publ., Singapore 2001 
\bibitem{Zuse}K.Zuse:``The Computing Universe'', Int.J.Theor.Phys. 21(1982)589
\bibitem{Feynman}R.P.Feynman:``Simulating Physics with Computers'', Int.J.Theor.Phys. 21(1982)467
\bibitem{Fredkin}E.Fredkin:``Digital Mechanics'', Physica D5(1990)254, see also the programmatic material on his homepage
\bibitem{Finkelstein}D.Finkelstein:``The Space-Time Code'', PR 184(1969)1261
\bibitem{Gardner}M.Gardner:``On CA, Self-Reproduction, the Garden of Eden and the Game of Life'', Sci.Am. 224(1971)112
\bibitem{Kauffman}S.Kauffman:``At Home in the Universe. The Search for Laws of Self-Organization and Complexity'', Oxford Univ.Pr., Oxford 1995
\bibitem{Langton}C.G.Langton ed.:``Artificial Life'', Proc. of an Interdisciplinary Workshop on the Synthesis and Silulation of Living Systems, Sept. 1987, Los Alamos, Addison-Wesley Publ., N.Y.
\bibitem{Waldrop}M.Waldrop:``Complexity, the Emerging Science at the Edge of Order and Chaos'', Penguin Books, N.Y. 1993
\bibitem{Nowotny1}T.Nowotny,M.Requardt:``Emergent Properties in Structurally Dynamic Disordered Cellular Networks'', J.Cell.Automata 2(2007)273, arXiv:cond-mat/0611427
\bibitem{Hooft4}G.'t Hooft:``Can Quantum Mechancs be Reconciled with CA?'', Int.J.Theor.Phys. 42
}(2003)349
\bibitem{Bollo1}B.Bollobas: ``Modern Graph Theory'', Springer, N.Y. 1998
\bibitem{Kauffman1}S.Kauffman: ``Origins of Order: Self-Organisation
  and Selection in Evolution'', Oxford Univ.Pr., Oxford 1993
\bibitem{Toffoli}T.Toffoli, N.Margolus: ``Cellular Automaton
  Machines'', MIT Pr., Cambridge Mass. 1987
\bibitem{Biggs}N.Biggs:``Algebraic Graph Theory'', 2nd edn., Cambridge Univ.Pr., Cambridge 1993
\bibitem{Godsil}C.Godsil,G.Royle:``Algebraic Graph Theory'', Springer, Berlin 2001
\bibitem{Dirac}M.Requardt:``Dirac Operators and the Calculation of the Connes Metric on Arbitrary (Infinite) Graphs'', J.Phys.A: Math.Gen. 35(2002)759, arXiv:math-ph/0108007
\bibitem{Susy}M.Requardt:``Supersymmetry on Graphs and Networks'', Int.J.Geom.Meth.Mod.Phys. 2(2005)585, arXiv:math-ph/0410059
\bibitem{Felsager}B.Felsager: ``Geometry, Particles, and Fields'',
  Springer, N.Y. 1998
\bibitem{Hodge}W.Hodge: ``The Theory and Applications of Harmonic
  Integrals'', 2nd ed., Cambridge Univ.Pr., Cambridge 1952
\bibitem{Edgar}G.A.Edgar: ``Measure, Topology, and Fractal Geometry'',
Springer, N.Y. 1990, or\\
K.Kuratowski: ``Topology'' Vol.1, Acad.Pr., N.Y. 1966
\bibitem{Harary}F.Buckley,F.Harary: ``Distance in Graphs'',
  Addison-Wesley, N.Y. 1990
\bibitem{Harpe}P.de la Harpe: ``Topics in Geometric Group Theory'',
  Univ. Chicago Pr., Chicago 2000
\bibitem{Falconer}K.J.Falconer: ``Fractal Geometry'', Wiley, Chichester 1990
\bibitem{Dhar}D.Dhar: ``Lattices of effectively nonintegral dimensionality'', J.Math.Phys. 18(1977)577
\bibitem{Watanabe}K.Hattori,T.Hattori,H.Watanabe: ``Gaussian Field Theories on General Networks and the Spectral Dimension'', Progr.Theor.Phys.Suppl. 92(1987)108
\bibitem{Filk}T.Filk: ``Equivalence of Massive Propagator Distance and Mathematical Distance on Graphs'', Mod.Phys.Lett. A 7(1992)2637
\bibitem{Durhuus}B.Durhuus: ``Hausdorff and Spectral Dimension of Infinite Random Graphs'', Act.Phys.Pol. B 40(2009)3509  
\bibitem{lumps}M.Requardt: ``(Quantum) Space-Time as a Statistical
  Geometry of Lumps in Random Networks'', Class.Quant.Grav. 17(2000)2029, gr-qc/9912059
\bibitem{Bollo2}B.Bollobas: ``Random Graphs'', Acad.Pr., N.Y.1985
\bibitem{Bollo3}B.Bollobas: ``Combinatorics'', Cambridge Univ.Pr.,
  London 1986
\bibitem{Roy}M.Requardt,S.Roy: ``(Quantum) Spacetime as a statistical geometry of fuzzy lumps and the connection with random metric spaces'', CQG 18(2001)3039, arXiv:gr-qc/9912059
\bibitem{Einstein}K.Menger in ``Albert Einstein:Philosopher Scientist, ed. P.A.Schilpp,n 3rd edn, Cambridge University Pr., London 1970
\bibitem{Pizana1}F.Larrion,V.Neumann-Lara,M.A.Pizana,T.D.Porter: ``Recognizing Self-Clique Graphs'', Mathematica Contemporanea 25(2003)125
\bibitem{Pizana2}M.A.Pizana: ``The Icosaheron is Clique Divergent'', Discr.Math. 262(2003)229
\bibitem{Pizana3}M.A.Pizana: ``Distances and Diameters on Iterated Clique Graphs'', Discr.Appl.Math. 141(2004)255
\bibitem{Gromov}M.Gromov: ``Metric Structures for Riemannian and Non-Riemannian Spaces'', Birkhaeuser, N.Y. 1998
\bibitem{Bridson}M.R.Bridson,A.Haeflinger: ``Metric Spaces of Non-Positive Curvature'', Springer, N.Y. 1999
\bibitem{Petersen}P.Petersen:``Gromov-Hausdorff Convergence of Metric
  Spaces'', AMS Proc.Pure Math. 54,3(1993)489
\bibitem{Gromov1}M.Gromov:``Groups of Polynomial Growth and Expanding
  Maps'', Publ.Math.IHES 53(1981)53
\bibitem{Petersen2}P.Petersen: ``Riemannian Geometry'', chapt.10,
  Springer, Berlin 1991
\bibitem{Quantum}M.Requardt: ``Quantum Theory as Emergent from an Undulatory Translocal Sub-Quantum Level'', arXiv:1205.1619
\bibitem{Watts}D.Watts: ``Small Worlds'', Princeton Univ.Pr., Princeton 1999
\bibitem{Strogatz}S.H.Strogatz: ``Nonlinear Dynamics and Chaos'', Perseus Books, Cambridge (USA) 1994
\bibitem{Requloch}A.Lochmann,M.Requardt: ``An Analysis of the
  Transition Zone Between the various Scaling Regimes in the
  Small-World Model'', J.Stat.Phys. 122(2006)255, arXiv:cond-mat/0409710
\bibitem{Strogatz2}S.H.Strogatz: ``From Kuramoto to Crawford: exploring the onset of synchronization in populations of coupled oscillators'', Physica D 143(2000)1
\bibitem{Niebur}E.Niebur,H.G.Schuster,D.M.Kammen,C.Koch: ``
Oscillator-phase coupling for different two-dimensional network connectivities'', PR A 44(1991)6895
\bibitem{Bohm}D.Bohm: ``Wholeness and the Implicate Order'', Routhledge and Kegan, London 1980

\end{thebibliography}
\end{document}